\documentclass[preprint,aps,superscriptaddress]{revtex4-2}

\usepackage{graphicx}
\usepackage{dcolumn}
\usepackage{bm}
\usepackage{amsthm}
\usepackage{amsmath}
\usepackage{amssymb}

\begin{document}

\title{Uniaxial Polarization Analysis of Bulk Ferromagnets:~Theory and First Experimental Results}

\author{Artem Malyeyev}\email{artem.malyeyev@uni.lu}
\author{Ivan Titov}
\address{Department of Physics and Materials Science, University of Luxembourg, 162A~Avenue de la Fa\"iencerie, L-1511~Luxembourg, Grand Duchy of Luxembourg}
\author{Charles D.\ Dewhurst}
\address{Institut Laue-Langevin, 6~Rue Jules Horowitz, B.P.~156, F-38042 Grenoble Cedex~9, France}
\author{Kiyonori Suzuki}
\address{Department of Materials Science and Engineering, Monash University, Clayton, Victoria~3800, Australia}
\author{Dirk Honecker}\email{dirk.honecker@stfc.ac.uk}
\address{ISIS Neutron and Muon Source, Rutherford Appleton Laboratory, Didcot,
United Kingdom}
\author{Andreas Michels}\email{andreas.michels@uni.lu}
\address{Department of Physics and Materials Science, University of Luxembourg, 162A~Avenue de la Fa\"iencerie, L-1511~Luxembourg, Grand Duchy of Luxembourg}

\begin{abstract}
Based on Brown's static equations of micromagnetics, we compute the uniaxial polarization of the scattered neutron beam of a bulk magnetic material. The theoretical expressions are compared to experimental data on a soft magnetic nanocrystalline alloy. The micromagnetic SANS theory provides a general framework for polarized real-space neutron methods, and it opens up a new avenue for magnetic neutron data analysis on magnetic microstructures.
\end{abstract}

\maketitle

\section{Introduction}

Polarized neutron scattering is one of the most powerful techniques for investigating the structure and dynamics of condensed matter, in particular magnetic materials and superconductors~\cite{tapan2006}. Based on the seminal papers by Bloch, Schwinger, and Halpern and Johnson~\cite{bloch1936,bloch1937,schwinger1937,halpern39}, the theory of polarized neutron scattering has been worked out in the early 1960's by Maleev and Blume~\cite{maleyev63,blume63}. Several classic experimental studies~\cite{shull51,moon69,rekveldt71,drabkin72,okorokov1978,pynn83,mezei86,schaerpf1993} have demonstrated the basic principles and paved the way for todays three-dimensional cryogenic polarization-analysis device (CRYOPAD)~\cite{tasset89,brown1993,tasset99,okorokov2001}. With this technique it becomes possible to measure 16~correlation functions, which provide important information on the nuclear and magnetic structure of materials (see Refs.~\onlinecite{williams,lovesey} for textbook expositions of polarized neutron scattering).

However, for the scattering of cold (long-wavelength) neutrons along the forward direction---as implemented on a small-angle neutron scattering (SANS) instrument---it has only in recent years become possible to perform `routinely' neutron-polarization analysis (retaining the full two-dimensional scattering information); more specifically, uniaxial (also called longitudinal or one dimensional) polarization analysis, where the polarization of the scattered neutrons is analyzed along the direction of the initial polarization~\cite{moon69}. Clearly, this progress is due to the development of efficient $^3$He spin filters (\textit{e.g.}\ \cite{heil2005,berna06,OKUDAIRA2020}), which, in contrast to \textit{e.g.}\ single-crystal analyzers, can be used over a rather broad wavelength range and cover a large detector acceptance angle. Note also that Refs.~\onlinecite{niketic2015,hautle2019,hautle2019a} report on the development of a novel neutron spin filter, based on the strong spin dependence of the neutron scattering on protons. For the combination of uniaxial polarization analysis with SANS, the term POLARIS has been coined~\cite{albi2005}. In contrast to CRYOPAD, which generally demands the sample to be in a zero magnetic field environment, POLARIS allows for the application of large magnetic fields.

The POLARIS method has been successfully employed for studying the superparamagnetic response of concentrated ferrofluids~\cite{albi2005}, proton domains in deuterated solutions~\cite{michels06a,aswal08nim,noda2016}, the multiferroic properties of $\mathrm{HoMn}_3$ single crystals~\cite{uehland2010}, the role of nanoscale heterogeneities for the magnetostriction of Fe-Ga alloys~\cite{laver2010a,laver2010b}, local weak ferromagnetism in $\mathrm{BiFeO}_3$~\cite{laver2011}, nanometer-sized magnetic domains and coherent magnetization reversal in an exchange-bias system~\cite{dufour2011}, precipitates in Heusler-based alloys~\cite{michelsheusler2019}, the magnetic microstructure of nanoscaled bulk magnets~\cite{michels2010epjb,michels2012prb2}, the internal spin structure of nanoparticles~\cite{kryckaprl2010,kryckaprl2014,ijiri2014,grutter2017,orue2018,bender2018jpcc,bender2018prb,oberdick2018,krycka2019,benderapl2019,honecker2020}, or Invar alloys~\cite{stewart2019}. Polarization analysis allows further to reveal the direction of the magnetic anisotropy in single-crystalline spin systems, \textit{e.g.}\ an easy plane versus an easy axis anisotropy or the confinement of the propagation vector along certain crystallographic directions in chiral and other exotic magnets~\cite{takagi2018,white2018}. In all of the above mentioned studies, the spin-resolved SANS cross sections were obtained and analyzed, but the polarization of the scattered neutrons was not further investigated.

Historically, this is the domain of the neutron depolarization technique~(see \textit{e.g.}\ \cite{holstein41b,hughes1948,hughes1949,burgy1950depol,maleevruban1970,rekveldt71,drabkin72,maleevruban1972,rekveldt73,rekveldt91} and references therein), where one measures the change in the polarization of a polarized neutron beam after transmission through a partially-magnetized magnetic material. Analysis of the $3 \times 3$ depolarization matrix yields information on \textit{e.g.}\ the average domain size and the domain magnetization. This type of polarization analysis on a SANS instrument has been termed `vector analysis of polarization' by Okorokov and Runov~\cite{okorokov2001}. Alternatively, it has been demonstrated that the neutron spin-echo technique can resolve magnetic small-angle scattering~\cite{grigosesans2006,rekveldt2006}. Spin-echo small-angle neutron scattering (SESANS) provides information on correlations on a length scale from about $10 \, \mathrm{nm}$ to $10 \, \mu\mathrm{m}$. In SESANS, the neutron spin precesses in a constant magnetic field and the neutron runtime difference due to sample scattering results in the dephasing of the neutron spins and in a loss of the measured beam polarization. Magnetic scattering can result in neutron spin-flip events that act as an additional optical element reversing the sense of the Larmor precession. The change of the neutron spin due to magnetic scattering can be exploited to study magnetic systems.

More recently, the method of polarized neutron dark-field contrast imaging (DFI) has been introduced for spatially-resolved small-angle scattering studies of magnetic microstructures~\cite{strobl2021}. First experimental results on a sintered Nd-Fe-B magnet demonstrated that not only dark-field contrast from half-polarized SANS measurements can be observed, but that it also becomes possible to separate and retrieve dark-field contrast for all spin-flip and non-spin flip channels separately. The polarized DFI method bears great potential for analyzing real-space magnetic correlations on a macroscopic length scale, well beyond of what can be probed with a conventional SANS instrument. Similarly, first measurements of micron-sized magnetic correlations have been performed with an alternative neutron precession technique called spin-echo modulated small-angle neutron scattering (SEMSANS)~\cite{parnell2021}. With this setup, the spin manipulations are performed before the sample so that the measurement is not sensitive to large stray fields (related \textit{e.g.}\ to the sample environment), and even allows studies under beam depolarizing conditions. The polarization analyzer discriminates the polarization parallel to the analyzing direction, such that the scattering cross sections for the opposite neutron spin state are probed at the sample. The two-dimensional neutron polarization modulation observed on the detector is then integrated to yield the one-dimensional correlation function of the system. 

In this work, we present a micromagnetic SANS theory for the uniaxial polarization of the scattered neutrons of bulk magnetic materials. The continuum theory of micromagnetics allows one to characterize the large-scale magnetization distribution of polycrystalline magnets, which is determined \textit{e.g.}\ by magnetic-anisotropy and saturation-magnetization fluctuations, antisymmetric exchange, and dipolar stray fields. Since the validity of micromagnetic theory extends to the micrometer regime, the here-developed theoretical framework may as well serve as a basis for the above-mentioned polarized neutron techniques (SESANS, DFI, SEMSANS). The derived theoretical expressions are tested against experimental SANS data on a soft magnetic nanocrystalline alloy.

The paper is organized as follows:~Section~\ref{poltheory} summarizes the elementary equations of polarized neutron scattering. Section~\ref{mumagtheory} sketches the basic steps and ideas of the micromagnetic SANS theory. Section~\ref{polscatt} displays the final expressions for the polarization of the scattered neutrons, discusses special sector averages, and shows the results for the $2\pi$-azimuthally-averaged saturated state. Section~\ref{expdet} furnishes the details of the polarized SANS experiment and on the investigated sample, while section~\ref{results} presents and discusses the analysis of the experimental results. Section~\ref{summary} summarizes the main findings of this study. Appendix~\ref{appa} features the expressions for the spin-resolved SANS cross sections in terms of the Fourier components of the magnetization, which enter the final expressions for the polarization, while Appendix~\ref{appb} showcases some computed examples for the polarization.

\section{Uniaxial SANS Polarization Analysis}
\label{poltheory}

\begin{figure}[tb!]
\centering
\resizebox{0.75\columnwidth}{!}{\includegraphics{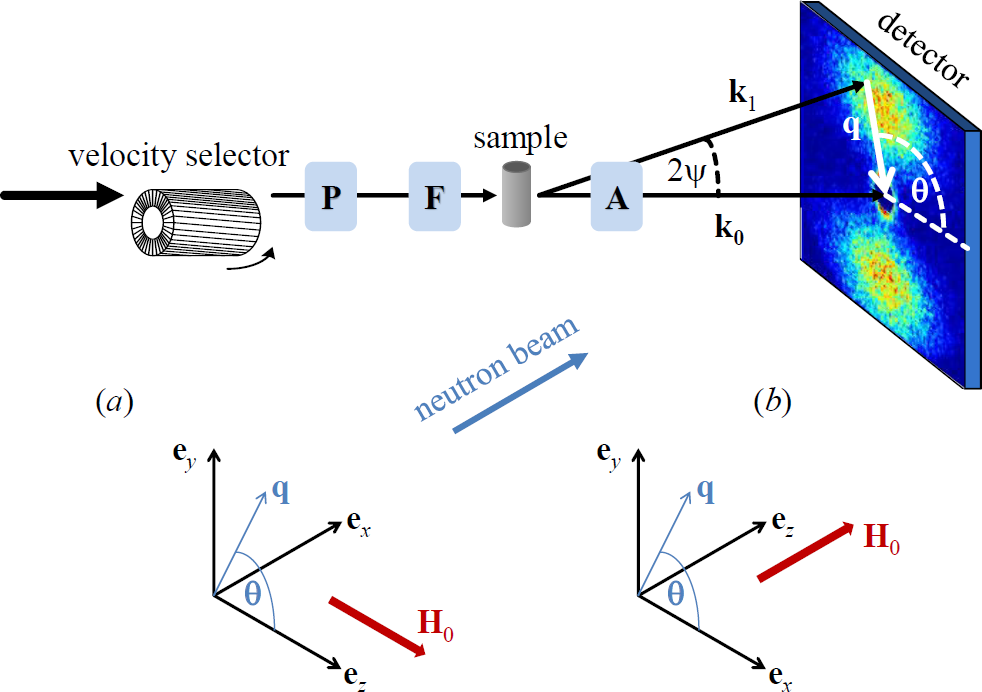}}
\caption{Sketch of the SANS setup and of the two most often employed scattering geometries in magnetic SANS experiments\index{scattering geometry}. (\textit{a})~Applied magnetic field $\mathbf{H}_0$ perpendicular to the incident neutron beam ($\mathbf{k}_0 \perp \mathbf{H}_0$); (\textit{b})~$\mathbf{k}_0 \parallel \mathbf{H}_0$. The momentum-transfer or scattering vector $\mathbf{q}$ corresponds to the difference between the wave vectors of the incident ($\mathbf{k}_0$) and the scattered ($\mathbf{k}_1$) neutrons, \textit{i.e.}\ $\mathbf{q} = \mathbf{k}_0 - \mathbf{k}_1$. Its magnitude for elastic scattering, $q = |\mathbf{q}| = (4\pi / \lambda) \sin(\psi)$, depends on the mean wavelength $\lambda$ of the neutrons and on the scattering angle $2\psi$. For a given $\lambda$, sample-to-detector distance $L_{\mathrm{SD}}$, and distance $r_{\mathrm{D}}$ from the center of the direct beam to a certain pixel element on the detector, the $q$-value can be obtained using $q \cong k_0 \frac{r_{\mathrm{D}}}{L_{\mathrm{SD}}}$. The symbols `P', `F', and `A' denote, respectively, the polarizer, spin flipper, and analyzer, which are optional neutron optical devices. Note that a second flipper after the sample has been omitted here. In spin-resolved SANS (POLARIS) using a $^3$He spin filter, the transmission (polarization) direction of the analyzer can be switched by $180^{\circ}$ by means of an rf pulse. SANS is usually implemented as elastic scattering ($k_0 = k_1 = 2\pi / \lambda$), and the component of $\mathbf{q}$ along the incident neutron beam [\textit{i.e.}\ $q_x$ in (\textit{a}) and $q_z$ in (\textit{b})] is neglected. The angle $\theta$ may be conveniently used in order to describe the angular anisotropy of the recorded scattering pattern on a two-dimensional position-sensitive detector. Image taken from Ref.~\onlinecite{michelsbook}.}
\label{fig1}
\end{figure}

Fig.~\ref{fig1} depicts a sketch of a typical uniaxial neutron polarization analysis setup. We consider the most relevant cases where the externally applied magnetic field $\mathbf{H}_0$, which defines the polarization axis for both the incident and scattered neutrons, is either perpendicular or parallel to the wave vector $\mathbf{k}_0$ of the incident neutron beam. Note that in \textit{both} scattering geometries $\mathbf{H}_0$ is assumed to be parallel to the $\mathbf{e}_z$-direction of a Cartesian laboratory coordinate system.

In a classical picture, the polarization $\mathbf{P}$ of a neutron beam containing $N$ spins can be defined as the average over the individual polarizations $\mathbf{P}_j$ of the neutrons as~\cite{schweizer2006}:
\begin{eqnarray}
\label{poldefschweizer}
\mathbf{P} = \frac{1}{N} \sum^{N}_{j=1} \mathbf{P}_j ,
\end{eqnarray}
where $0 \leq |\mathbf{P}| \leq 1$. In experimental SANS studies the beam is usually partially polarized along a certain guide-field direction (quantization axis), which we take here as the $z$-direction. Assuming that the expectation values of the perpendicular polarization components vanish, \textit{i.e.}\ $P_x = P_y = 0$, and that $P_z = P$, one can then introduce the fractions
\begin{eqnarray}
\label{polfractions}
p^+ = \frac{1}{2} \left( 1 + P \right) \hspace{0.5cm} \mathrm{and} \hspace{0.5cm} p^- = \frac{1}{2} \left( 1 - P \right)
\end{eqnarray}
of neutrons in the spin-up ($+$) and spin-down ($-$) state, with 
\begin{eqnarray}
\label{polfractionsconditions}
p^+ + p^- = 1 \hspace{0.5cm} \mathrm{and} \hspace{0.5cm} p^+ - p^- = P .
\end{eqnarray}
Obviously, for an unpolarized beam $p^+ = p^- = 0.5$ and $P = 0$, while $P = + 1$ ($p^+ = 1$) or $P = - 1$ ($p^- = 1$) for a fully polarized beam.

When there is an additional analyzer behind the sample, configured such that it selects only neutrons with spins either parallel or antiparallel to the initial polarization, then one can distinguish four scattering cross sections (scattering processes)~\cite{blume63,moon69,schweizer2006}:~two of which conserve the neutron-spin direction ($++$ and $--$), the so-called non-spin-flip cross sections
\begin{eqnarray}
\label{nsfchapterone}
\frac{d \Sigma^{++}}{d \Omega} &=& K \left[ b^{-2}_{\mathrm{H}} |\widetilde{N}|^2 + b^{-1}_{\mathrm{H}} (\widetilde{N} \widetilde{Q}_z^{\ast} + \widetilde{N}^{\ast} \widetilde{Q}_z) + |\widetilde{Q}_z|^2 \right] , \nonumber \\
\frac{d \Sigma^{--}}{d \Omega} &=& K \left[ b^{-2}_{\mathrm{H}} |\widetilde{N}|^2 - b^{-1}_{\mathrm{H}} (\widetilde{N} \widetilde{Q}_z^{\ast} + \widetilde{N}^{\ast} \widetilde{Q}_z) + |\widetilde{Q}_z|^2 \right] , \nonumber \\
\end{eqnarray}
and two cross sections which reverse the neutron spin ($+-$ and $-+$), the so-called spin-flip cross sections
\begin{eqnarray}
\label{sfchapterone}
\frac{d \Sigma^{+-}}{d \Omega} &=& K \left[ |\widetilde{Q}_x|^2 + |\widetilde{Q}_y|^2 - i (\widetilde{Q}_x \widetilde{Q}_y^{\ast} - \widetilde{Q}_x^{\ast} \widetilde{Q}_y) \right] , \nonumber \\
\frac{d \Sigma^{-+}}{d \Omega} &=& K \left[ |\widetilde{Q}_x|^2 + |\widetilde{Q}_y|^2 + i (\widetilde{Q}_x \widetilde{Q}_y^{\ast} - \widetilde{Q}_x^{\ast} \widetilde{Q}_y) \right] . \nonumber \\
\end{eqnarray}
In these expressions, $K = 8 \pi^3 V^{-1} b_{\mathrm{H}}^2$, where $V$ denotes the scattering volume, and $b_{\mathrm{H}} = 2.70 \times 10^{-15} \, \mathrm{m} \, \mu^{-1}_{\mathrm{B}} = 2.91 \times 10^{8} \, \mathrm{A^{-1} m^{-1}}$ is a constant (with $\mu_{\mathrm{B}}$ the Bohr magneton), which relates the atomic magnetic moment $\mu_{\mathrm{a}}$ to the atomic magnetic scattering length $b_{\mathrm{m}}$, given by~\cite{moon69}:
\begin{eqnarray}
\label{bmagdef}
b_{\mathrm{m}} = \frac{\gamma_{\mathrm{n}} r_0}{2} \frac{\mu_{\mathrm{a}}}{\mu_{\mathrm{B}}} f(\mathbf{q}) \cong 2.70 \times 10^{-15} \, \mathrm{m} \frac{\mu_\mathrm{a}}{\mu_\mathrm{B}} f(\mathbf{q}) \cong b_{\mathrm{H}} \mu_{\mathrm{a}} ,
\end{eqnarray}
where $\gamma_{\mathrm{n}} = 1.913$ denotes the neutron magnetic moment expressed in units of the nuclear magneton, $r_0 = 2.818 \times 10^{-15} \, \mathrm{m}$ is the classical radius of the electron, and $f(\mathbf{q})$ is the normalized atomic magnetic form factor, which we set to unity, $f \cong 1$, along the forward direction. The function $\widetilde{N}(\mathbf{q})$ denotes the Fourier transform of the nuclear scattering-length density $N(\mathbf{r})$. The partial SANS cross sections, equations~(\ref{nsfchapterone}) and (\ref{sfchapterone}), are written here in terms of the Cartesian components of the Halpern--Johnson vector $\widetilde{\mathbf{Q}}$ (sometimes also denoted as the magnetic interaction or magnetic scattering vector)~\cite{halpern39}:
\begin{eqnarray}
\label{hjvector}
\widetilde{\mathbf{Q}} = \hat{\mathbf{q}} \times \left( \hat{\mathbf{q}} \times \widetilde{\mathbf{M}} \right) = \hat{\mathbf{q}} \left( \hat{\mathbf{q}} \cdot  \widetilde{\mathbf{M}} \right) - \widetilde{\mathbf{M}} ,
\end{eqnarray}
where $\hat{\mathbf{q}}$ is the unit scattering vector, and $\widetilde{\mathbf{M}}(\mathbf{q}) = \{ \widetilde{M}_x(\mathbf{q}), \widetilde{M}_y(\mathbf{q}), \widetilde{M}_z(\mathbf{q}) \}$ represents the Fourier transform of the magnetization vector field $\mathbf{M}(\mathbf{r}) = \{ M_x(\mathbf{r}), M_y(\mathbf{r}), M_z(\mathbf{r}) \}$ of the sample under study. The three-dimensional Fourier-transform pair of the magnetization is defined as follows:
\begin{eqnarray}
\label{fouriertrafodef}
\widetilde{\mathbf{M}}(\mathbf{q}) = \frac{1}{(2\pi)^{3/2}} \int\limits_{-\infty}^{+\infty} \int\limits_{-\infty}^{+\infty} \int\limits_{-\infty}^{+\infty} \mathbf{M}(\mathbf{r}) \exp\left( -i \mathbf{q} \cdot \mathbf{r} \right) d^3r , \\
\mathbf{M}(\mathbf{r}) = \frac{1}{(2\pi)^{3/2}} \int\limits_{-\infty}^{+\infty} \int\limits_{-\infty}^{+\infty} \int\limits_{-\infty}^{+\infty} \widetilde{\mathbf{M}}(\mathbf{q}) \exp\left( i \mathbf{q} \cdot \mathbf{r} \right) d^3q .
\end{eqnarray}
The Halpern--Johnson vector is a manifestation of the dipolar origin of magnetic neutron scattering and it emphasizes the fact that only the components of $\mathbf{M}$ which are perpendicular to $\mathbf{q}$ are relevant for magnetic neutron scattering. We note that different symbols for the Halpern--Johnson vector such as $\mathbf{M}_{\perp}$, $\mathbf{Q}_{\perp}$, $\mathbf{S}_{\perp}$, or $\mathbf{q}$, as in the original paper by Halpern and Johnson~\cite{halpern39}, can be found in the literature. Likewise, in many textbooks (\textit{e.g.}\ \cite{lovesey,squires}) $\widetilde{\mathbf{Q}}$ is defined with a minus sign and normalized by the factor $2 \mu_{\mathrm{B}}$, which makes it dimensionless. $\widetilde{\mathbf{Q}}$ is a linear vector function of the components of $\widetilde{\mathbf{M}}$. Both $\widetilde{\mathbf{Q}}(\mathbf{q})$ and $\widetilde{\mathbf{M}}(\mathbf{q})$ are in general complex vectors. For $\mathbf{k}_0 \perp \mathbf{H}_0$ and $\mathbf{k}_0 \parallel \mathbf{H}_0$ one finds, respectively (subscripts $\perp$ and $\parallel$ refer to the respective scattering geometry, compare Fig.~\ref{fig1}):
\begin{eqnarray}
\label{qperpdef}
\hat{\mathbf{q}}_{\perp} &=& \{0, \sin\theta, \cos\theta \} , \\
\label{qparadef}
\hat{\mathbf{q}}_{\parallel} &=& \{ \cos\theta, \sin\theta, 0 \} .
\end{eqnarray}
Inserting these expressions into equation~(\ref{hjvector}) yields:
\begin{eqnarray}
\label{hjvectorperp}
\widetilde{\mathbf{Q}}_{\perp} = \left\{ \begin{array}{c} 
- \widetilde{M}_x \\ 
- \widetilde{M}_y \cos^2\theta + \widetilde{M}_z \sin\theta \cos\theta \\ 
  \widetilde{M}_y \sin\theta \cos\theta - \widetilde{M}_z \sin^2\theta \end{array} \right\} ,
\end{eqnarray}
\begin{eqnarray}
\label{hjvectorpara}
\widetilde{\mathbf{Q}}_{\parallel} = \left\{ \begin{array}{c} 
- \widetilde{M}_x \sin^2\theta + \widetilde{M}_y \sin\theta \cos\theta \\ 
  \widetilde{M}_x \sin\theta \cos\theta - \widetilde{M}_y \cos^2\theta \\ 
- \widetilde{M}_z \end{array} \right\} .
\end{eqnarray}
Inspection of equations~(\ref{nsfchapterone}) and (\ref{sfchapterone}) shows that the transversal components $\widetilde{Q}_x$ and $\widetilde{Q}_y$ give rise to spin-flip scattering, while the longitudinal component $\widetilde{Q}_z$ results in non-spin-flip scattering. Furthermore, if we set $\theta = 0^{\circ}$ in equation~(\ref{hjvectorperp}), which corresponds to the case that the scattering vector is along the neutron polarization, we see that
\begin{eqnarray}
\label{hjvectorperpthetanull}
\widetilde{\mathbf{Q}}_{\perp}^{\theta = 0^{\circ}} = \left\{ \begin{array}{c} 
- \widetilde{M}_x \\ 
- \widetilde{M}_y \\ 
  0 \end{array} \right\} ,
\end{eqnarray}
so that nuclear coherent and magnetic scattering are fully separated in the perpendicular scattering geometry. In the case $\mathbf{k}_0 \parallel \mathbf{H}_0$ [equation~(\ref{hjvectorpara})], spin-flip scattering probes only the transversal magnetization Fourier components $\widetilde{M}_{x,y}$, whereas the longitudinal scattering is entirely contained in the non-spin-flip channel, in contrast to the $\mathbf{k}_0 \perp \mathbf{H}_0$ geometry. We emphasize that nuclear-spin-dependent SANS is not taken into account in this paper, so that the corresponding scattering contributions do not show up in equations~(\ref{nsfchapterone}) and (\ref{sfchapterone}).

The total SANS cross section $d \Sigma / d \Omega$ can be expressed in terms of the initial spin populations $p^{\pm}$ as~\cite{blume63,moon69,schweizer2006}:
\begin{eqnarray}
\label{elasticmagneticcrossgeneralcontchapteronepolarized1expressionone}
\frac{d \Sigma}{d \Omega} = p^+ \frac{d \Sigma^{++}}{d \Omega} + p^+ \frac{d \Sigma^{+-}}{d \Omega} + p^- \frac{d \Sigma^{--}}{d \Omega} + p^- \frac{d \Sigma^{-+}}{d \Omega} . \nonumber \\
\end{eqnarray}
Inserting the above expressions for $p^+$ and $p^-$ [equations~(\ref{polfractions}) and (\ref{polfractionsconditions})] and for the partial SANS cross sections $d \Sigma^{\pm \pm} /d \Omega$ and $d \Sigma^{\pm \mp} /d \Omega$ [equations~(\ref{nsfchapterone}) and (\ref{sfchapterone})], equation~(\ref{elasticmagneticcrossgeneralcontchapteronepolarized1expressionone}) evaluates to:
\begin{widetext}
\begin{eqnarray}
\label{elasticmagneticcrossgeneralcontchapteronepolarized1}
\frac{d \Sigma}{d \Omega} = K \left[ b^{-2}_{\mathrm{H}} |\widetilde{N}|^2 + |\widetilde{\mathbf{Q}}|^2 + \mathbf{P} \cdot b^{-1}_{\mathrm{H}} (\widetilde{N} \widetilde{\mathbf{Q}}^{\ast} + \widetilde{N}^{\ast} \widetilde{\mathbf{Q}}) - i \mathbf{P} \cdot (\widetilde{\mathbf{Q}} \times \widetilde{\mathbf{Q}}^{\ast}) \right] ,
\end{eqnarray}
which, using $\mathbf{P} = \{ 0, 0, P_z = \pm P \}$, can be rewritten as:
\begin{eqnarray}
\label{elasticmagneticcrossgeneralcontchapteronepolarized2}
\frac{d \Sigma^{\pm}}{d \Omega} = K \left[ b^{-2}_{\mathrm{H}} |\widetilde{N}|^2 + |\widetilde{\mathbf{Q}}|^2 \pm P b^{-1}_{\mathrm{H}} (\widetilde{N} \widetilde{Q}_z^{\ast} + \widetilde{N}^{\ast} \widetilde{Q}_z) \mp i P ( \widetilde{Q}_x \widetilde{Q}_y^{\ast} - \widetilde{Q}_x^{\ast} \widetilde{Q}_y) \right] . \,\,\,\,\,\,
\end{eqnarray}
\end{widetext}
Since the cross section is a scalar quantity and the polarization is an axial vector (or pseudovector), equation~(\ref{elasticmagneticcrossgeneralcontchapteronepolarized1}) shows that the system under study must itself contain an axial vector. As emphasized by Maleev~\cite{maleyev2002}, examples for such built-in pseudovectors are related to the interaction of a polycrystalline sample with an external magnetic field (inducing an average magnetization directed along the applied field), the existence of a spontaneous magnetization in a ferromagnetic single crystal, the antisymmetric Dzyaloshinskii--Moriya interaction (DMI), mechanical (torsional) deformation, or the presence of spin spirals. If on the other hand there is no preferred axis in the system, then $d \Sigma / d \Omega$ is independent of $\mathbf{P}$. Examples include a collection of randomly oriented noninteracting nuclear (electronic) spins, which describe the general case of nuclear (paramagnetic) scattering at not-too-low temperatures and large applied fields, or a multi-domain ferromagnet with a random distribution of the domains. The same condition---existence of an axial system vector---applies for neutrons to be polarized in the scattering process [compare the last two terms in equation~(\ref{finalpolarizationcrosssection}) below].

In the domain of magnetic SANS it is customary to denote experiments with a polarized incident beam only, and no spin analysis of the scattered neutrons, with the acronym SANSPOL. The two SANSPOL cross sections $d \Sigma^{+} / d \Omega$ and $d \Sigma^{-} / d \Omega$ (also sometimes denoted as the half-polarized SANS cross sections) combine non-spin-flip and spin-flip scattering contributions, according to ($p^{\pm} = 1$):
\begin{eqnarray}
\label{sumcrosssanspolchapter1a}
\frac{d \Sigma^{+}}{d \Omega} &=& \frac{d \Sigma^{++}}{d \Omega} + \frac{d \Sigma^{+-}}{d \Omega} , \\
\label{sumcrosssanspolchapter1b}
\frac{d \Sigma^{-}}{d \Omega} &=&
\frac{d \Sigma^{--}}{d \Omega} + \frac{d \Sigma^{-+}}{d \Omega} .
\end{eqnarray}
Finally, noting that an unpolarized beam can be viewed as consisting of $50 \%$ spin-up and $50 \%$ spin-down neutrons [compare equations~(\ref{polfractions}) and (\ref{polfractionsconditions})], the unpolarized SANS cross section is obtained as [compare equation~(\ref{elasticmagneticcrossgeneralcontchapteronepolarized1expressionone})]:
\begin{eqnarray}
\label{sumcrossall}
\frac{d \Sigma}{d \Omega} &=& \frac{1}{2} \left( \frac{d \Sigma^{++}}{d \Omega} + \frac{d \Sigma^{+-}}{d \Omega} + \frac{d \Sigma^{--}}{d \Omega} + \frac{d \Sigma^{-+}}{d \Omega}\right) \nonumber \\ 
&=& \frac{1}{2} \left( \frac{d \Sigma^{+}}{d \Omega} + \frac{d \Sigma^{-}}{d \Omega} \right) .
\end{eqnarray}
The polarization $P_{\mathrm{f}}$ of the scattered beam along the direction of the incident neutron polarization $P$ is obtained from the following relation~\cite{blume63,moon69,schweizer2006}:
\begin{eqnarray}
\label{finalpolarizationcrosssection}
P_{\mathrm{f}} \frac{d \Sigma}{d \Omega} &=& p^+ \frac{d \Sigma^{++}}{d \Omega} + p^- \frac{d \Sigma^{-+}}{d \Omega} - p^- \frac{d \Sigma^{--}}{d \Omega} - p^+ \frac{d \Sigma^{+-}}{d \Omega} \nonumber \\
 &=& K P \left[ b^{-2}_{\mathrm{H}} |\widetilde{N}|^2 + |\widetilde{Q}_z|^2 - |\widetilde{Q}_x|^2 - |\widetilde{Q}_y|^2 \right] \nonumber \\
&+& K \left[ b^{-1}_{\mathrm{H}} (\widetilde{N} \widetilde{Q}_z^{\ast} + \widetilde{N}^{\ast} \widetilde{Q}_z) + i (\widetilde{Q}_x \widetilde{Q}_y^{\ast} - \widetilde{Q}_x^{\ast} \widetilde{Q}_y) \right] .
\end{eqnarray}
The first four terms in the second line on the right-hand side of equation~(\ref{finalpolarizationcrosssection}) demonstrate that nuclear scattering (to be more precise, the nuclear coherent scattering, the isotopic disorder scattering, and $1/3$ of the nuclear-spin-dependent scattering) and scattering due to the longitudinal component $\widetilde{Q}_z$ of the magnetic scattering vector $\widetilde{\mathbf{Q}}$ do not reverse the initial polarization, while the two transversal components $\widetilde{Q}_x$ and $\widetilde{Q}_y$ give rise to spin-flip scattering. The last two terms in equation~(\ref{finalpolarizationcrosssection}) do create polarization:~these are the familiar nuclear-magnetic interference terms ($\widetilde{N} \widetilde{Q}_z^{\ast} + \widetilde{N}^{\ast} \widetilde{Q}_z$), which are commonly used to polarize beams, and the chiral term $i (\widetilde{Q}_x \widetilde{Q}_y^{\ast} - \widetilde{Q}_x^{\ast} \widetilde{Q}_y)$, which is of relevance in inelastic scattering (dynamic chirality)~\cite{oko81,maleyev2002,grigo2015}, in elastic scattering on spiral structures and weak ferromagnets (canted antiferromagnets)~\cite{hutanu2021}, or in the presence of the DMI in microstructural-defect-rich magnets~\cite{michelsdmi2019,quan2020}. We remind the reader that nuclear-spin-dependent scattering is not taken into account in the expressions for the magnetic SANS cross sections. In the general expression for the polarization of the scattered neutrons, a term $i \mathbf{P} \times (\widetilde{N} \widetilde{\mathbf{Q}}^{\ast} - \widetilde{N}^{\ast} \widetilde{\mathbf{Q}})$ appears~\cite{schweizer2006}, which is ignored in equation~(\ref{finalpolarizationcrosssection}). This term rotates the polarization perpendicular to the initial polarization and cannot be observed in the uniaxial setup. We emphasize that in linear neutron polarimetry it is not possible to distinguish between a rotation of the polarization vector and a change of its length~\cite{moon69,maleyev2002}.

From equation~(\ref{finalpolarizationcrosssection}) it follows that the polarization $P_{\mathrm{f}}(\mathbf{q})$ of the scattered neutron beam at momentum-transfer vector $\mathbf{q}$ can be expressed as~\cite{maleyev63,blume63,brown2006}:
\begin{eqnarray}
\label{polgen}
\label{finalpolarizationcrosssectionfinal}
P_{\mathrm{f}} &=& \frac{p^+ \frac{d \Sigma^{++}}{d \Omega} + p^- \frac{d \Sigma^{-+}}{d \Omega} - p^- \frac{d \Sigma^{--}}{d \Omega} - p^+ \frac{d \Sigma^{+-}}{d \Omega}}{p^+ \frac{d \Sigma^{++}}{d \Omega} + p^+ \frac{d \Sigma^{+-}}{d \Omega} + p^- \frac{d \Sigma^{--}}{d \Omega} + p^- \frac{d \Sigma^{-+}}{d \Omega}} ,
\end{eqnarray}
which for $p^+ = 1$ ($p^- = 0$) and $p^- = 1$ ($p^+ = 0$) reduce to, respectively:
\begin{subequations}
\begin{eqnarray}
\label{finalpolarizationcrosssectionplus}
P_{\mathrm{f}}^{+} = \frac{\frac{d \Sigma^{++}}{d \Omega} - \frac{d \Sigma^{+-}}{d \Omega}}{\frac{d \Sigma^{++}}{d \Omega} + \frac{d \Sigma^{+-}}{d \Omega}} = 1 - 2 \frac{\frac{d \Sigma^{+-}}{d \Omega}}{\frac{d \Sigma^{+}}{d \Omega}} , \\
\label{finalpolarizationcrosssectionminus}
P_{\mathrm{f}}^{-} = \frac{\frac{d \Sigma^{-+}}{d \Omega} - \frac{d \Sigma^{--}}{d \Omega}}{\frac{d \Sigma^{--}}{d \Omega} + \frac{d \Sigma^{-+}}{d \Omega}} = - \left( 1 - 2 \frac{\frac{d \Sigma^{-+}}{d \Omega}}{\frac{d \Sigma^{-}}{d \Omega}} \right).
\end{eqnarray}
\end{subequations}
Note that for the following analysis, we drop the minus sign in front of the round brackets in equation~(\ref{finalpolarizationcrosssectionminus}). For a quantitative analysis of $P_{\mathrm{f}}^{\pm}$, a theoretical model for the magnetization Fourier components $\widetilde{M}_{x,y,z}(\mathbf{q})$ and for $\widetilde{N}(\mathbf{q})$ is required. This will be discussed in the next section.

\section{Sketch of Micromagnetic SANS Theory}
\label{mumagtheory}

In Ref.~\onlinecite{michelsPRB2016} we have presented a theory for the magnetic SANS cross section of bulk ferromagnets. The approach, which considers the response of the magnetization to spatially inhomogeneous magnetic anisotropy fields and magnetostatic fields, is based on the continuum theory of micromagnetics, valid close to magnetic saturation, and takes the antisymmetric DMI into account. Here, we recall the basic steps and ideas. The starting point are the static equations of micromagnetics for the bulk, which can be conveniently written as~\cite{brown,aharonibook,kronfahn03}:
\begin{equation}
\label{torque}
\mathbf{M}(\mathbf{r}) \times \mathbf{H}_{\mathrm{eff}}(\mathbf{r}) = 0 .
\end{equation}
Equations~(\ref{torque}) express the fact that at static equilibrium the torque on the magnetization $\mathbf{M}(\mathbf{r})$ due to an effective magnetic field $\mathbf{H}_{\mathrm{eff}}(\mathbf{r})$ vanishes everywhere inside the material. The effective field is defined as the functional derivative of the ferromagnet's total energy-density functional $\epsilon$ with respect to the magnetization:
\begin{eqnarray}
\label{heff}
\mathbf{H}_{\mathrm{eff}} &=& - \frac{1}{\mu_0} \frac{\delta \epsilon}{\delta \mathbf{M}} \nonumber \\
&=& \mathbf{H}_0 + \mathbf{H}_{\mathrm{d}} + \mathbf{H}_{\mathrm{p}} + \mathbf{H}_{\mathrm{ex}} + \mathbf{H}_{\mathrm{DMI}} ,
\end{eqnarray}
where $\mathbf{H}_0$ is a uniform applied magnetic field, $\mathbf{H}_{\mathrm{d}}(\mathbf{r})$ is the magnetostatic field, $\mathbf{H}_{\mathrm{p}}(\mathbf{r})$ is the magnetic anisotropy field, $\mathbf{H}_{\mathrm{ex}} = l_{\mathrm{M}}^2 \nabla^2 \mathbf{M}$ is the exchange field, and $\mathbf{H}_{\mathrm{DMI}} = - l_{\mathrm{D}} \nabla \times \mathbf{M}$ is due to the DMI (assuming for simplicity a cubic symmetry); $\nabla^2$ is the Laplace operator and $\nabla = \partial/\partial x \, \mathbf{e}_x + \partial/\partial y \, \mathbf{e}_y + \partial/\partial z \, \mathbf{e}_z$ is the gradient operator, where the unit vectors along the Cartesian laboratory axes are, respectively, denoted with $\mathbf{e}_x$, $\mathbf{e}_y$, and $\mathbf{e}_z$ ($\mu_0$:~vacuum permeability). The micromagnetic length scales
\begin{equation}
\label{lmmmdef}
l_{\mathrm{M}} = \sqrt{\frac{2 A}{\mu_0 M_0^2}}
\end{equation}
and
\begin{equation}
\label{lddddef}
l_{\mathrm{D}} = \frac{2 D}{\mu_0 M_0^2}
\end{equation}
are, respectively, related to the magnetostatic interaction and to the DMI. The values for the DMI constant $D$ and for the exchange-stiffness constant $A$ are assumed to be uniform throughout the material, in contrast to the local saturation magnetization $M_{\mathrm{s}}(\mathbf{r})$, which is assumed to depend explicitly on the position $\mathbf{r}$ (see also Ref.~\onlinecite{metmi2015}); $M_0 = V^{-1} \int_V M_{\mathrm{s}}(\mathbf{r}) dV$ denotes the macroscopic saturation magnetization of the sample, which can be measured with a magnetometer. Typical values for the above length scales are $l_{\mathrm{M}} \cong 5-10 \, \mathrm{nm}$~\cite{kronfahn03} and $l_{\mathrm{D}} \cong 1-2 \, \mathrm{nm}$. However, due to the lack of an established database for $D$-values, the latter estimate should be considered with some care.

The solution of the linearized version of equation~(\ref{torque}) (see Ref.~\onlinecite{michelsPRB2016} for details) provides closed-form expressions for the transversal Fourier components $\widetilde{M}_x(\mathbf{q})$ and $\widetilde{M}_y(\mathbf{q})$. Together with theoretical models (or even experimental data) for the longitudinal magnetization Fourier component $\widetilde{M}_z(\mathbf{q})$ and for the nuclear scattering amplitude $\widetilde{N}(\mathbf{q})$, these analytical solutions can be employed to compute the corresponding terms in the SANS cross sections (see Appendix~\ref{appa}) and, hence, to evaluate the final polarizations. Averaging over the directions of the magnetic anisotropy field in the plane perpendicular to the applied field, the magnetic terms for the transversal magnetic field geometry ($\mathbf{k}_0 \perp \mathbf{H}_0$, $q_x = 0$) are:
\begin{widetext}
\begin{equation}
\label{mxp}
|\widetilde{M}_x|^2 = \frac{p^2}{2} \frac{\widetilde{H}^2_{\mathrm{p}} \left( \left[ 1 + p \sin^2\theta \right]^2 + p^2 l_{\mathrm{D}}^2 q^2 \cos^2\theta \right) + 2 \widetilde{M}_z^2 (1 + p)^2 l_{\mathrm{D}}^2 q^2 \sin^2\theta}{\left( 1 + p \sin^2\theta - p^2 l_{\mathrm{D}}^2 q^2 \cos^2\theta \right)^2} , 
\end{equation}
\begin{equation}
\label{myp}
|\widetilde{M}_y|^2 = \frac{p^2}{2} \frac{\widetilde{H}_{\mathrm{p}}^2 \left( 1 + p^2 l_{\mathrm{D}}^2 q^2 \cos^2\theta \right) + 2 \widetilde{M}_z^2 \left( 1 + p l_{\mathrm{D}}^2 q^2 \right)^2 \sin^2\theta \cos^2\theta}{\left( 1 + p \sin^2\theta - p^2 l_{\mathrm{D}}^2 q^2 \cos^2\theta \right)^2} , 
\end{equation}
\begin{equation}
\label{mymzp}
CT_{yz} = \widetilde{M}_y \widetilde{M}_z^{\ast} + \widetilde{M}_y^{\ast} \widetilde{M}_z = - \frac{2 \widetilde{M}_z^2 p \left( 1 + p l_{\mathrm{D}}^2 q^2 \right) \sin\theta \cos\theta}{1 + p \sin^2\theta - p^2 l_{\mathrm{D}}^2 q^2 \cos^2\theta} .
\end{equation}
\begin{equation}
\label{ffinal}
- 2 i \chi = \frac{2 \widetilde{H}^2_{\mathrm{p}} p^3 \left( 2 + p \sin^2\theta \right) l_{\mathrm{D}} q \cos^3\theta + 4 \widetilde{M}_z^2 p (1 + p)^2 l_{\mathrm{D}} q \sin^2\theta \cos\theta}{\left( 1 + p \sin^2\theta - p^2 l_{\mathrm{D}}^2 q^2 \cos^2\theta \right)^2} ,
\end{equation}
The results for the parallel field geometry ($\mathbf{k}_0 \parallel \mathbf{H}_0$, $q_z = 0$) are:
\begin{equation}
\label{mxpara}
|\widetilde{M}_x|^2 = \frac{p^2}{2} \frac{\widetilde{H}_{\mathrm{p}}^2 \left( 1 + p (2 + p) \sin^2\theta \right) +
2 \widetilde{M}_z^2 (1 + p)^2 l_{\mathrm{D}}^2 q^2 \sin^2\theta}{\left( 1 + p \right)^2} ,
\end{equation}
\begin{equation}
\label{mypara}
|\widetilde{M}_y|^2 = \frac{p^2}{2} \frac{\widetilde{H}_{\mathrm{p}}^2 \left( 1 + p (2 + p) \cos^2\theta \right) + 
2 \widetilde{M}_z^2 (1 + p)^2 l_{\mathrm{D}}^2 q^2 \cos^2\theta}{\left( 1 + p \right)^2} ,
\end{equation}
\begin{equation}
\label{mxmypara}
CT_{xy} = \widetilde{M}_x \widetilde{M}_y^{\ast} + \widetilde{M}_x^{\ast} \widetilde{M}_y = - p^2 \frac{\left( \widetilde{H}_{\mathrm{p}}^2 p (2 + p) + 2 \widetilde{M}_z^2 (1 + p)^2 l_{\mathrm{D}}^2 q^2 \right) \sin\theta \cos\theta}{\left( 1 + p \right)^2} ,
\end{equation}
\begin{equation}
\label{ffinalpara}
\chi = 0 .
\end{equation}
\end{widetext}
In equations~(\ref{mxp})$-$(\ref{ffinalpara}), $\widetilde{H}^2_{\mathrm{p}}(q \xi_{\mathrm{H}})$ denotes the magnitude-square of the Fourier transform of the magnetic anisotropy field, and $\widetilde{M}^2_z(q \xi_{\mathrm{M}})$ is the Fourier component of the longitudinal magnetization. We also emphasize that
\begin{equation}
\label{msfteqmzft}
\widetilde{M}_z \cong \widetilde{M}_{\mathrm{s}}
\end{equation}
is assumed in the approach-to-saturation regime. These functions characterize the strength and spatial structure of, respectively, the magnetic anisotropy field $\mathbf{H}_{\mathrm{p}}(\mathbf{r})$, with correlation length $\xi_{\mathrm{H}}$, and of the local saturation magnetization $M_{\mathrm{s}}(\mathbf{r})$, with correlation length $\xi_{\mathrm{M}}$.
\begin{equation}
\label{pdef}
p(q, H_{\mathrm{i}}) = \frac{M_0}{H_{\mathrm{eff}}(q, H_{\mathrm{i}})} = \frac{M_0}{H_{\mathrm{i}} \left( 1 + l_{\mathrm{H}}^2 q^2 \right)}
\end{equation}
is a dimensionless function, where 
\begin{equation}
\label{heffdef}
H_{\mathrm{eff}}(q, H_{\mathrm{i}}) = H_{\mathrm{i}} \left( 1 + l_{\mathrm{H}}^2 q^2 \right) 
\end{equation}
is the effective magnetic field [not to be confused with $\mathbf{H}_{\mathrm{eff}}(\mathbf{r})$ in equation~(\ref{torque})], which depends on the internal magnetic field $H_{\mathrm{i}} = H_0 - N M_0$ ($N$:~demagnetizing factor), on $q = |\mathbf{q}|$, and on the exchange length of the field
\begin{equation}
\label{lhdef}
l_{\mathrm{H}} = \sqrt{\frac{2A}{\mu_0 M_0 H_{\mathrm{i}}}} .
\end{equation}
The latter quantity is a measure for the size of inhomogeneously magnetized regions around microstructural lattice defects~\cite{mettus2015}. The Fourier coefficient of the longitudinal magnetization $\widetilde{M}_z(\mathbf{q})$ provides information on the spatial variation of the saturation magnetization $M_{\mathrm{s}}(\mathbf{r})$; for instance, $|\widetilde{M}_z|^2 \propto (\Delta M)^2$ in a multiphase magnetic nanocomposite, where $\Delta M$ denotes the jump of the magnetization magnitude at internal (particle-matrix) interfaces~\cite{michels2013}. Moreover, while the squared Fourier components and the cross terms are even functions of $\mathbf{q}$, it is easily seen that the chiral term $-2i \chi(\mathbf{q})$ [equation~(\ref{ffinal})] is asymmetric in $\mathbf{q}$, which is due to the DMI term:~at small fields, when the term $\propto \widetilde{H}^2_{\mathrm{p}}$ in the numerator of equation~(\ref{ffinal}) dominates, two extrema parallel and antiparallel to the field axis are observed, whereas at larger fields, when the term $\propto \widetilde{M}_z^2$ dominates, additional maxima and minima appear approximately along the detector diagonals~\cite{michelsPRB2016}.

By inserting equations~(\ref{mxp})$-$(\ref{ffinalpara}) into the SANS cross sections (see Appendix~\ref{appa}) and by specifying particular models for the nuclear scattering function $\widetilde{N}^2(q \xi_{\mathrm{nuc}})$, the longitudinal magnetic Fourier component $\widetilde{M}^2_z(q \xi_{\mathrm{M}})$, and for the Fourier coefficient of the magnetic anisotropy field $\widetilde{H}^2_{\mathrm{p}}(q \xi_{\mathrm{H}})$, one obtains $P^{\pm}$ as a function of the magnitude and orientation of the scattering vector $\mathbf{q}$, the applied magnetic field $\mathbf{H}_0$, the magnetic-interaction parameters ($A$, $D$, $M_0$, $\Delta M$, $H_{\mathrm{p}}$, $\xi_{\mathrm{M}}$, $\xi_{\mathrm{H}}$), and microstructural quantities (particle-size distribution, crystallograpic texture, etc.). Later on in the paper, for graphically displaying the $P^{\pm}$, we have assumed that $\widetilde{N}^2$, $\widetilde{M}_z^2$, and $\widetilde{H}_{\mathrm{p}}^2$ are all isotropic (\textit{i.e.}\ $\theta$-independent), as is appropriate for polycrystalline texture-free bulk ferromagnets, and that they can be represented by Lorentzian-squared functions, \textit{i.e.}\
\begin{equation}
\label{mzsquaredmodel}
\widetilde{M}^2_z(q \xi_{\mathrm{M}}) = \frac{A_{\mathrm{M}}^2 \xi_{\mathrm{M}}^6}{\left( 1 + q^2 \xi_{\mathrm{M}}^2 \right)^2} ,
\end{equation}
\begin{equation}
\label{hanisquaredmodel}
\widetilde{H}_{\mathrm{p}}^2(q \xi_{\mathrm{H}}) = \frac{A_{\mathrm{H}}^2 \xi_{\mathrm{H}}^6}{\left( 1 + q^2 \xi_{\mathrm{H}}^2 \right)^2} ,
\end{equation}
\begin{equation}
\label{nucmodel}
\widetilde{N}^2(q \xi_{\mathrm{nuc}}) = \alpha(q) b_{\mathrm{H}}^2 \widetilde{M}^2_z(q \xi_{\mathrm{M}}) ,
\end{equation}
where the amplitudes $A_{\mathrm{M}}^2$ and $A_{\mathrm{H}}^2$ (both in units of $\mathrm{A^2}/\mathrm{nm^2}$) are, respectively, related to the mean-square magnetization fluctuation and anisotropy-field variation. Of course, other scattering functions such as the form factor of a sphere and various structure-factor models (\textit{e.g.}\ a Percus-Yevick hard-sphere structure factor) can be straightforwardly implemented~\cite{mettus2015}. The characteristic structure sizes of $\widetilde{M}_z^2$ and $\widetilde{H}_{\mathrm{p}}^2$ are generally different. We remind the reader that these are related, respectively, to the spatial extent of regions with uniform saturation magnetization ($\xi_{\mathrm{M}}$) and magnetic anisotropy field ($\xi_{\mathrm{H}}$). Measurement of the magnetic field dependent Guinier radius provides a means to determine these correlation lengths as well as the exchange-stiffness constant~$A$~\cite{michelsmgl2020}. A simple example where $\xi_{\mathrm{M}} = \xi_{\mathrm{H}}$ is a collection of homogeneous and defect-free magnetic nanoparticles in a magnetic and homogeneous matrix. If, on the other hand, atomically-sharp grain boundaries are introduced into such particles, then the direction of the magnetic anisotropy field changes due to the changing set of crystallographic directions at the intraparticle interfaces, but the value of $M_{\mathrm{s}}$ may remain the same, so that $\xi_{\mathrm{H}} < \xi_{\mathrm{M}}$. In Ref.~\onlinecite{michels2013} it was shown, assuming $\xi_{\mathrm{H}} = \xi_{\mathrm{M}}$ and using the sphere form factor for both $\widetilde{M}^2_z$ and $\widetilde{H}_{\mathrm{p}}^2$, that it is the ratio of $H_{\mathrm{p}} / \Delta M$ (related to the amplitudes $A_{\mathrm{H}}$ and $A_{\mathrm{M}}$) which determines the angular anisotropy and the asymptotic power-law dependence of $d \Sigma / d \Omega$ as well as the characteristic decay length of spin-misalignment fluctuations. The ratio of nuclear to longitudinal magnetic scattering is denoted with $\alpha$, which for the general case (that the nuclear correlation length $\xi_{\mathrm{nuc}}$ is different from $\xi_{\mathrm{M}}$) is a function of $q$. Here, we do not specify a particular $\xi_{\mathrm{nuc}}$ and assume this characteristic size to be contained in $\alpha(q)$.

\section{Polarization of Scattered Beam}
\label{polscatt}

When the expressions for the elastic differential spin-flip and SANSPOL cross sections $\frac{d \Sigma^{\pm \mp}}{d \Omega}$ and $\frac{d \Sigma^{\pm}}{d \Omega}$ [equations~(\ref{sfperp})$-$(\ref{sanspolpara})] are inserted into equations~(\ref{finalpolarizationcrosssectionplus}) and (\ref{finalpolarizationcrosssectionminus}), and use is made of the expressions for the magnetization Fourier components [equations~(\ref{mxp})$-$(\ref{ffinalpara})], one obtains, respectively, for the transversal and longitudinal scattering geometry:
\begin{widetext}
\begin{subequations}
\begin{eqnarray}
\centering
\label{polgen1}
P_{\mathrm{f}\perp}^{+}(\mathbf{q}) &=& 1 - 2 \frac{h_1(\mathbf{q}) - i \chi(\mathbf{q})}{h_2(\mathbf{q}) - b_{\mathrm{H}}^{-1} CT_{\widetilde{N} \widetilde{M}_z} \sin^2\theta + b_{\mathrm{H}}^{-1} CT_{\widetilde{N} \widetilde{M}_y} \sin\theta \cos\theta - i \, \chi(\mathbf{q})} , \\
\label{polgen2}
P_{\mathrm{f}\perp}^{-}(\mathbf{q}) &=& 1 - 2 \frac{h_1(\mathbf{q}) + i \chi(\mathbf{q})}{h_2(\mathbf{q}) + b_{\mathrm{H}}^{-1} CT_{\widetilde{N} \widetilde{M}_z} \sin^2\theta - b_{\mathrm{H}}^{-1} CT_{\widetilde{N} \widetilde{M}_y} \sin\theta \cos\theta + i \, \chi(\mathbf{q})} ,
\end{eqnarray}
\begin{eqnarray}
\centering
\label{polgen3}
P_{\mathrm{f}\parallel}^{+}(\mathbf{q}) &=& 1 - 2 \frac{g_1(\mathbf{q})}{g_2(\mathbf{q}) - b_{\mathrm{H}}^{-1} CT_{\widetilde{N} \widetilde{M}_z}} , \\
\label{polgen4}
P_{\mathrm{f}\parallel}^{-}(\mathbf{q}) &=& 1 - 2 \frac{g_1(\mathbf{q})}{g_2(\mathbf{q}) + b_{\mathrm{H}}^{-1} CT_{\widetilde{N} \widetilde{M}_z}} .
\end{eqnarray}
\end{subequations}
\end{widetext}
The functions $h_1(\mathbf{q})$, $h_2(\mathbf{q})$, $g_1(\mathbf{q})$, and $g_2(\mathbf{q})$ are \textit{independent} of the incident neutron beam polarization and are defined as:
\begin{eqnarray}
\label{h1vonq}
h_1(\mathbf{q}) = |\widetilde{M}_x|^2 + |\widetilde{M}_y|^2 \cos^4\theta \nonumber \\ + |\widetilde{M}_z|^2 \sin^2\theta \cos^2\theta - CT_{yz} \sin\theta \cos^3\theta ,
\end{eqnarray}
\begin{eqnarray}
\label{h2vonq}
h_2(\mathbf{q}) = b_{\mathrm{H}}^{-2} |\widetilde{N}|^2 + |\widetilde{M}_x|^2 + |\widetilde{M}_y|^2 \cos^2\theta \nonumber \\ + |\widetilde{M}_z|^2 \sin^2\theta - CT_{yz} \sin\theta \cos\theta ,
\end{eqnarray}
\begin{eqnarray}
\label{g1vonq}
g_1(\mathbf{q}) = |\widetilde{M}_x|^2 \sin^2\theta + |\widetilde{M}_y|^2 \cos^2\theta \nonumber \\ - CT_{xy} \sin\theta \cos\theta ,
\end{eqnarray}
\begin{eqnarray}
\label{g2vonq}
g_2(\mathbf{q}) = b_{\mathrm{H}}^{-2} |\widetilde{N}|^2 + |\widetilde{M}_x|^2 \sin^2\theta + |\widetilde{M}_y|^2 \cos^2\theta \nonumber \\ + |\widetilde{M}_z|^2 - CT_{xy} \sin\theta \cos\theta .
\end{eqnarray}
At complete magnetic saturation, when $\mathbf{M}(\mathbf{r}) = \{ 0, 0, M_{\mathrm{s}}(\mathbf{r}) \}$, these expressions reduce to:
\begin{eqnarray}
\label{h1vonqsat}
h^{\mathrm{sat}}_1(\mathbf{q}) = |\widetilde{M}_{\mathrm{s}}|^2 \sin^2\theta \cos^2\theta ,
\end{eqnarray}
\begin{eqnarray}
\label{h2vonqsat}
h^{\mathrm{sat}}_2(\mathbf{q}) = b_{\mathrm{H}}^{-2} |\widetilde{N}|^2 + |\widetilde{M}_{\mathrm{s}}|^2 \sin^2\theta ,
\end{eqnarray}
\begin{eqnarray}
\label{g1vonqsat}
g^{\mathrm{sat}}_1(\mathbf{q}) = 0 ,
\end{eqnarray}
\begin{eqnarray}
\label{g2vonqsat}
g^{\mathrm{sat}}_2(\mathbf{q}) = b_{\mathrm{H}}^{-2} |\widetilde{N}|^2 + |\widetilde{M}_{\mathrm{s}}|^2 ,
\end{eqnarray}
where $\widetilde{M}_{\mathrm{s}}(\mathbf{q})$ is the Fourier transform of $M_{\mathrm{s}}(\mathbf{r})$. As can be seen from equations~(\ref{polgen1}) and (\ref{polgen2}), the difference between $P_{\mathrm{f}}^{+}$ and $P_{\mathrm{f}}^{-}$ resides, for $\mathbf{k}_0 \perp \mathbf{H}_0$, in the nuclear-magnetic interference terms $\propto \widetilde{N} \widetilde{M}_z$ and $\propto \widetilde{N} \widetilde{M}_y$, and in $\chi(\mathbf{q})$, while for $\mathbf{k}_0 \parallel \mathbf{H}_0$ both polarizations differ only by the term $\propto \widetilde{N} \widetilde{M}_z$ [equations~(\ref{polgen3}) and (\ref{polgen4})]. We would also like to remind the reader that the Fourier coefficients in the above expressions are evaluated in the plane of the detector, which for the perpendicular scattering geometry corresponds to the plane $q_x = 0$, and to the plane $q_z = 0$ for the parallel geometry (compare Fig.~\ref{fig1}). 

The $\widetilde{N} \widetilde{M}_y$ contribution to equations~(\ref{polgen1}) and (\ref{polgen2}) requires a special consideration. This term is expected to be negligible for a polycrystalline statistically-isotropic ferromagnet with vanishing fluctuations of the saturation magnetization. This can be seen by scrutinizing the following expression for the $\widetilde{M}_y$~magnetization Fourier component in the perpendicular scattering geometry, corresponding to the plane $q_x = 0$~\cite{michelsbook}:
\begin{equation}
\label{solmyqx0book}
\widetilde{M}_y = \frac{p \left( \widetilde{H}_{\mathrm{p}y} - \widetilde{M}_z \frac{q_y q_z}{q^2} \left( 1 + p l_{\mathrm{D}}^2 q^2 \right) - i \widetilde{H}_{\mathrm{p}x} p l_{\mathrm{D}} q_z \right)}{1 + p \frac{q_y^2}{q^2} - p^2 l_{\mathrm{D}}^2 q_z^2} ,
\end{equation}
where $\widetilde{H}_{\mathrm{p}x}$ and $\widetilde{H}_{\mathrm{p}y}$ denote the Cartesian components of the Fourier transform of the magnetic anisotroy field. If we assume that the nuclear scattering is isotropic and that $\widetilde{H}_{\mathrm{p}x}$ and $\widetilde{H}_{\mathrm{p}y}$ vary randomly in the plane perpendicular to the field, then the corresponding averages over the direction of the anisotropy field vanish. The only remaining term in the $\widetilde{N} \widetilde{M}_y$ contribution is then ($q_y/q = \sin\theta$; $q_z/q = \cos\theta$):
\begin{eqnarray}
\label{nmyct}
CT_{\widetilde{N} \widetilde{M}_y} &=& \widetilde{N} \widetilde{M}_y^{\ast} + \widetilde{N}^{\ast} \widetilde{M}_y \nonumber \\ 
&=& - \frac{2 \widetilde{N} \widetilde{M}_z p \left( 1 + p l_{\mathrm{D}}^2 q^2 \right) \sin\theta \cos\theta}{1 + p \sin^2\theta - p^2 l_{\mathrm{D}}^2 q^2 \cos^2\theta} ,
\end{eqnarray}
where we have furthermore assumed that both $\widetilde{N}(q)$ and $\widetilde{M}_z(q)$ are real-valued, as is done throughout this paper. Note that equation~(\ref{nmyct}) still needs to be multiplied with the term $\sin\theta \cos\theta$ in order to obtain the corresponding contribution to $P_{\mathrm{f}\perp}^{\pm}$ [compare equations~(\ref{polgen1}) and (\ref{polgen2})]. For homogeneous single-phase materials with $M_{\mathrm{s}}=$~constant, the $\widetilde{N} \widetilde{M}_y$ contribution is expected to be negligible, and we are not aware that is has been reported experimentally. However, for materials exhibiting strong spatial nanoscale variations in the saturation magnetization, \textit{i.e.}\ $M_{\mathrm{s}} = M_{\mathrm{s}}(\mathbf{r})$, such as magnetic nanocomposites or porous ferromagnets, it should be possible to observe this scattering contribution in polarized SANS experiments.

\subsection{Sector Averages}

Carrying out a $2\pi$~azimuthal average of the $P_{\mathrm{f}}^{\pm}(\mathbf{q})$, which are maps with numbers varying between $\pm 1$, may result in a significant loss of information (compare \textit{e.g.}\ Figs.~\ref{figa1} and \ref{figa2} below). It is therefore often advantageous to consider cuts of $P_{\mathrm{f}}^{\pm}$ along certain directions in $\mathbf{q}$-space. This might also be of relevance for other spin-manipulating techniques such as SEMSANS, which is a one dimensional technique that only measures correlations in the encoding direction~\cite{parnell2021}. Sector averages are straightforwardly obtained by evaluating equations~(\ref{polgen1})$-$(\ref{polgen4}) [using equations~(\ref{h1vonq})$-$(\ref{g2vonqsat})] for certain angles $\theta$. For instance, for the perpendicular scattering geometry and for $\mathbf{q}$ along the vertical direction on the detector ($\theta = 90^{\circ}$), we obtain [$\chi(q, \theta = 90^{\circ}) = 0$, compare equation~(\ref{ffinal})]:
\begin{widetext}
\begin{subequations}
\begin{eqnarray}
\centering
\label{polgen1t90}
P_{\mathrm{f}\perp}^{+}(q, \theta = 90^{\circ}) &=& 1 - 2 \frac{|\widetilde{M}_x|^2}{b_{\mathrm{H}}^{-2} |\widetilde{N}|^2 + |\widetilde{M}_x|^2 + |\widetilde{M}_z|^2 - b_{\mathrm{H}}^{-1} CT_{\widetilde{N} \widetilde{M}_z}} , \\
\label{polgen2t90}
P_{\mathrm{f}\perp}^{-}(q, \theta = 90^{\circ}) &=& 1 - 2 \frac{|\widetilde{M}_x|^2}{b_{\mathrm{H}}^{-2} |\widetilde{N}|^2 + |\widetilde{M}_x|^2 + |\widetilde{M}_z|^2 + b_{\mathrm{H}}^{-1} CT_{\widetilde{N} \widetilde{M}_z}} ,
\end{eqnarray}
\end{subequations}
\end{widetext}
where [compare equation~(\ref{mxp})]
\begin{equation}
\label{mxpt90}
|\widetilde{M}_x|^2(q, \theta = 90^{\circ}) = \frac{p^2}{2} \left( \widetilde{H}^2_{\mathrm{p}}  + 2 \widetilde{M}_z^2 l_{\mathrm{D}}^2 q^2 \right) .
\end{equation}
At saturation ($M_x = 0$), $P_{\mathrm{f}\perp}^{\pm}(q, \theta = 90^{\circ}) = 1$, except for the case $\alpha = 1$, where $P_{\mathrm{f}\perp}^{+}(q, \theta = 90^{\circ}) = -1$. We also see that information on the DMI is contained in $|\widetilde{M}_x|^2$ via the length scale $l_{\mathrm{D}}$. For $l_{\mathrm{D}} = 0$, $|\widetilde{M}_x|^2 = \frac{p^2}{2} \widetilde{H}^2_{\mathrm{p}}$.

For the perpendicular scattering geometry and $\mathbf{q}$ along the horizontal direction ($\theta = 0^{\circ}$), we obtain:
\begin{subequations}
\begin{eqnarray}
\centering
\label{polgen1t0}
P_{\mathrm{f}\perp}^{+}(q, \theta = 0^{\circ}) &=& 1 - 2 \frac{|\widetilde{M}_x|^2 + |\widetilde{M}_y|^2 - i \chi}{b_{\mathrm{H}}^{-2} |\widetilde{N}|^2 + |\widetilde{M}_x|^2 + |\widetilde{M}_y|^2 - i \chi} \nonumber , \\ \\
\label{polgen2t0}
P_{\mathrm{f}\perp}^{-}(q, \theta = 0^{\circ}) &=& 1 - 2 \frac{|\widetilde{M}_x|^2 + |\widetilde{M}_y|^2 + i \chi}{b_{\mathrm{H}}^{-2} |\widetilde{N}|^2 + |\widetilde{M}_x|^2 + |\widetilde{M}_y|^2 + i \chi} \nonumber , \\
\end{eqnarray}
\end{subequations}
where [compare equations~(\ref{mxp})$-$(\ref{myp})]
\begin{equation}
\label{mxpt0}
|\widetilde{M}_x|^2(q, \theta = 0^{\circ}) = |\widetilde{M}_y|^2 = \frac{p^2}{2} \frac{\widetilde{H}^2_{\mathrm{p}} \left( 1 + p^2 l_{\mathrm{D}}^2 q^2 \right)}{\left( 1 - p^2 l_{\mathrm{D}}^2 q^2 \right)^2}
\end{equation}
and [compare equation~(\ref{ffinal})]
\begin{equation}
\label{chit0}
i \chi(q, \theta = 0^{\circ}) = - \frac{2 \widetilde{H}^2_{\mathrm{p}} p^3 l_{\mathrm{D}} q}{\left( 1 - p^2 l_{\mathrm{D}}^2 q^2 \right)^2} .
\end{equation}
At saturation ($M_x = M_y = \chi = 0$) and for nonzero nuclear scattering, $P_{\mathrm{f}\perp}^{\pm}(q, \theta = 0^{\circ}) = 1$. For $l_{\mathrm{D}} = 0$, $|\widetilde{M}_x|^2 + |\widetilde{M}_y|^2 = p^2 \widetilde{H}^2_{\mathrm{p}}$ and $P_{\mathrm{f}\perp}^{\pm}(q, \theta = 0^{\circ})$ contains information on the transversal spin components.

\subsection{Saturated State}

At saturation and for $\mathbf{k}_0 \perp \mathbf{H}_0$, it is readily verified from equations~(\ref{polgen1}) and (\ref{polgen2}) using equations~(\ref{h1vonqsat})$-$(\ref{g2vonqsat}), $\chi(\mathbf{q}) = 0$, and $CT_{\widetilde{N} \widetilde{M}_y} = 0$ that
\begin{subequations}
\begin{eqnarray}
\label{psat1}
P_{\mathrm{f}\perp}^{+}(q,\theta) = 1 - 2 \frac{\sin^2\theta \cos^2\theta}{\alpha - 2 \sqrt{\alpha} \sin^2\theta + \sin^2\theta} \\
\label{psat2}
P_{\mathrm{f}\perp}^{-}(q,\theta) = 1 - 2 \frac{\sin^2\theta \cos^2\theta}{\alpha + 2 \sqrt{\alpha} \sin^2\theta + \sin^2\theta}
\end{eqnarray}
\end{subequations}
depend exclusively on the ratio
\begin{equation}
\label{alphadef}
\alpha(q) = \frac{\widetilde{N}^2(q)}{b_{\mathrm{H}}^2 \widetilde{M}^2_s(q)}
\end{equation}
of nuclear to longitudinal magnetic scattering [compare equation~(\ref{nucmodel})]. The possible angular anisotropy of $\alpha$ is not considered in this paper. Since for $\mathbf{k}_0 \parallel \mathbf{H}_0$ the spin-flip SANS cross section vanishes at saturation ($g_1 = 0$), we see that $P_{\mathrm{f}\parallel}^{\pm} = +1$. The azimuthally-averaged ($\frac{1}{2\pi} \int (\cdots) d\theta$) versions of equations~(\ref{psat1}) and (\ref{psat2}) read:
\begin{subequations}
\begin{eqnarray}
\label{psataziplus}
P_{\mathrm{f}\perp}^{+}(q) &=& \frac{2 \sqrt{\alpha} \left( -1 + \sqrt{\alpha} + \left| -1 +\sqrt{\alpha} \right| \right)}{\left( 1 - 2 \sqrt{\alpha} \right)^2} , \\
\label{psataziminus}
P_{\mathrm{f}\perp}^{-}(q) &=& 1 - \frac{1}{\left( 1 + 2 \sqrt{\alpha} \right)^2} .
\end{eqnarray}
\end{subequations}
The function $\alpha(q)$ can be a monotonically increasing or decreasing function of $q$, and it can even exhibit local extrema. In the following, we will consider the cases of $\alpha =$~constant and $\alpha = \alpha(q)$ using the experimental data of the soft magnetic Fe-based alloy NANOPERM~\cite{michels2012prb2}.

\subsubsection{$\alpha =$~constant}

\begin{figure}[tb!]
\centering
\resizebox{0.90\columnwidth}{!}{\includegraphics{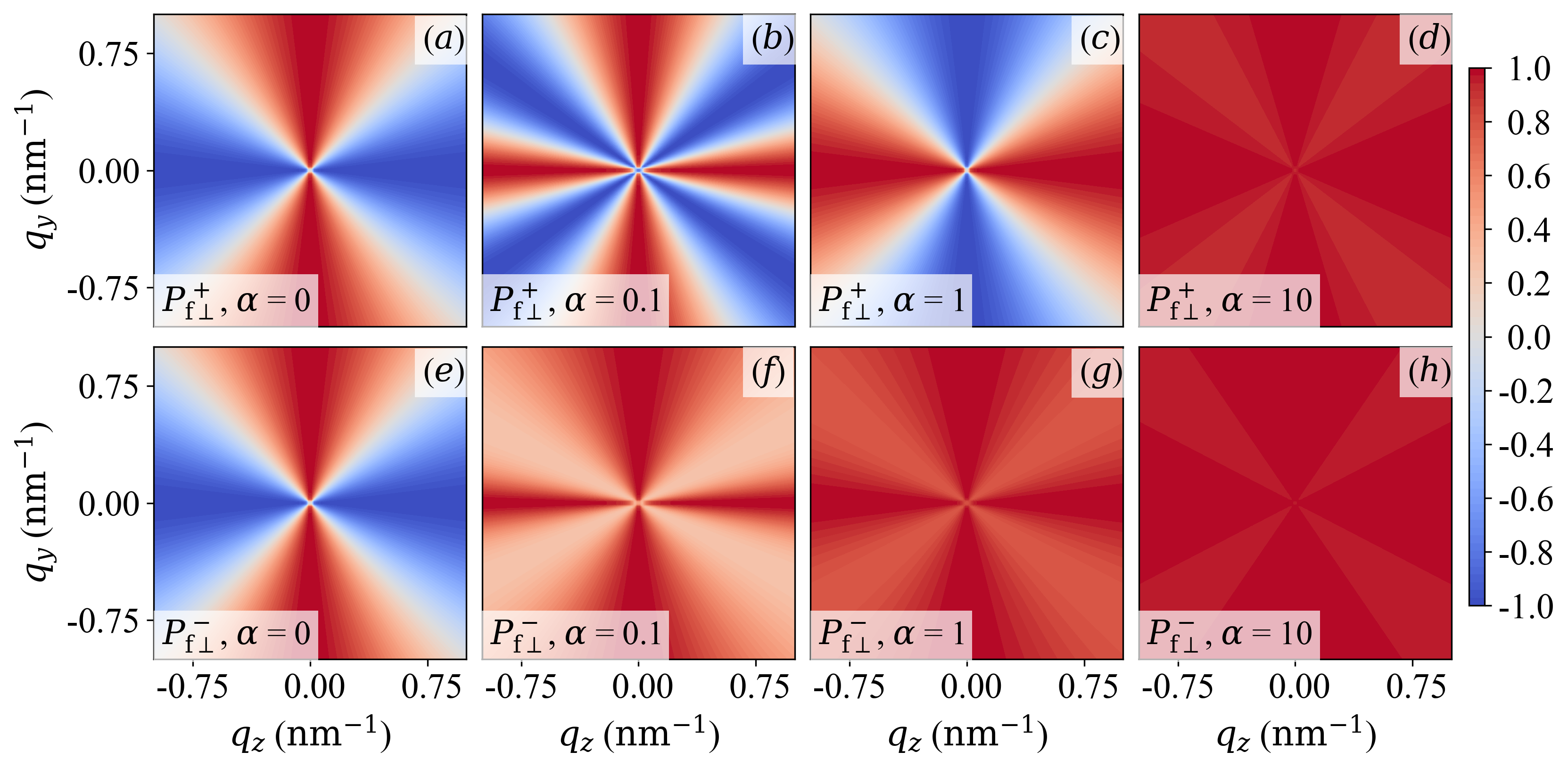}}
\caption{Plot of $P_{\mathrm{f}\perp}^{+}(q_y, q_z)$ (upper row) and $P_{\mathrm{f}\perp}^{-}(q_y, q_z)$ (lower row) in the saturate state for different values of $\alpha$ (see insets) [equations~(\ref{psat1}) and (\ref{psat2})].}
\label{fig2}
\end{figure}

\begin{figure}[tb!]
\centering
\resizebox{0.70\columnwidth}{!}{\includegraphics{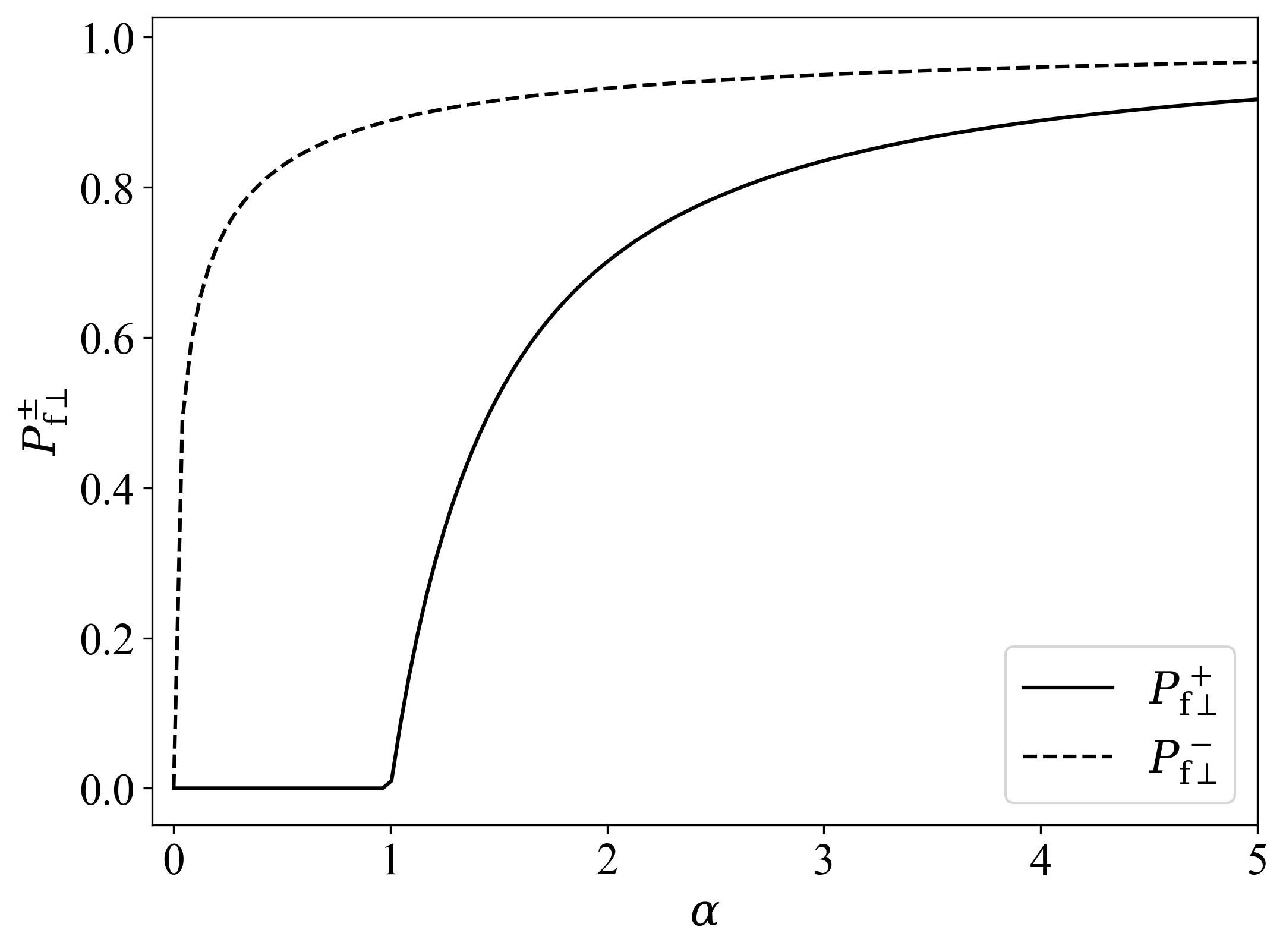}}
\caption{Plot of $P_{\mathrm{f}\perp}^{+}$ and $P_{\mathrm{f}\perp}^{-}$ (see inset) in the saturate state as a function of $\alpha$ [equations~(\ref{psataziplus}) and (\ref{psataziminus})].}
\label{fig3}
\end{figure}

Fig.~\ref{fig2} displays the two-dimensional polarization $P_{\mathrm{f}\perp}^{\pm}(\mathbf{q})$ of the scattered neutrons in the saturated state as a function of $\alpha =$~constant. The case of constant $\alpha$ is very rarely realized in experimental situations, and we consider it here only as a starting point for our discussion and for the comparison to the experimentally more relevant situation of $\alpha = \alpha(q)$. For $\alpha \rightarrow 0$ it follows that $P_{\mathrm{f}\perp}^{\pm} = 1 - 2 \cos^2\theta$ [Fig.~\ref{fig2}(\textit{a}) and (\textit{\textit{e}})], while $P_{\mathrm{f}\perp}^{+} = 1 - 2 \sin^2\theta$ [Fig.~\ref{fig2}(\textit{c})] and $P_{\mathrm{f}\perp}^{-} = 1 - 2 \sin^2\theta \cos^2\theta/(1 + 3 \sin^2\theta)$ [Fig.~\ref{fig2}(\textit{g})] for $\alpha = 1$. When nuclear coherent scattering is dominating ($\alpha \rightarrow \infty$), we see that the $P_{\mathrm{f}\perp}^{\pm}$ both tend to unity, as expected. The corresponding $2\pi$-azimuthally-averaged functions [equations~(\ref{psataziplus}) and (\ref{psataziminus})] are plotted in Fig.~\ref{fig3}. One readily verifies that $P_{\mathrm{f}\perp}^{+} = 0$ for $\sqrt{\alpha} \leq 1$, which further underlines the loss of information when a $2\pi$~azimuthal average is carried out [compare \textit{e.g.}\ Fig.~\ref{fig2}(\textit{b})].

\subsubsection{$\alpha = \alpha(q)$}

\begin{figure}[tb!]
\centering
\resizebox{0.70\columnwidth}{!}{\includegraphics{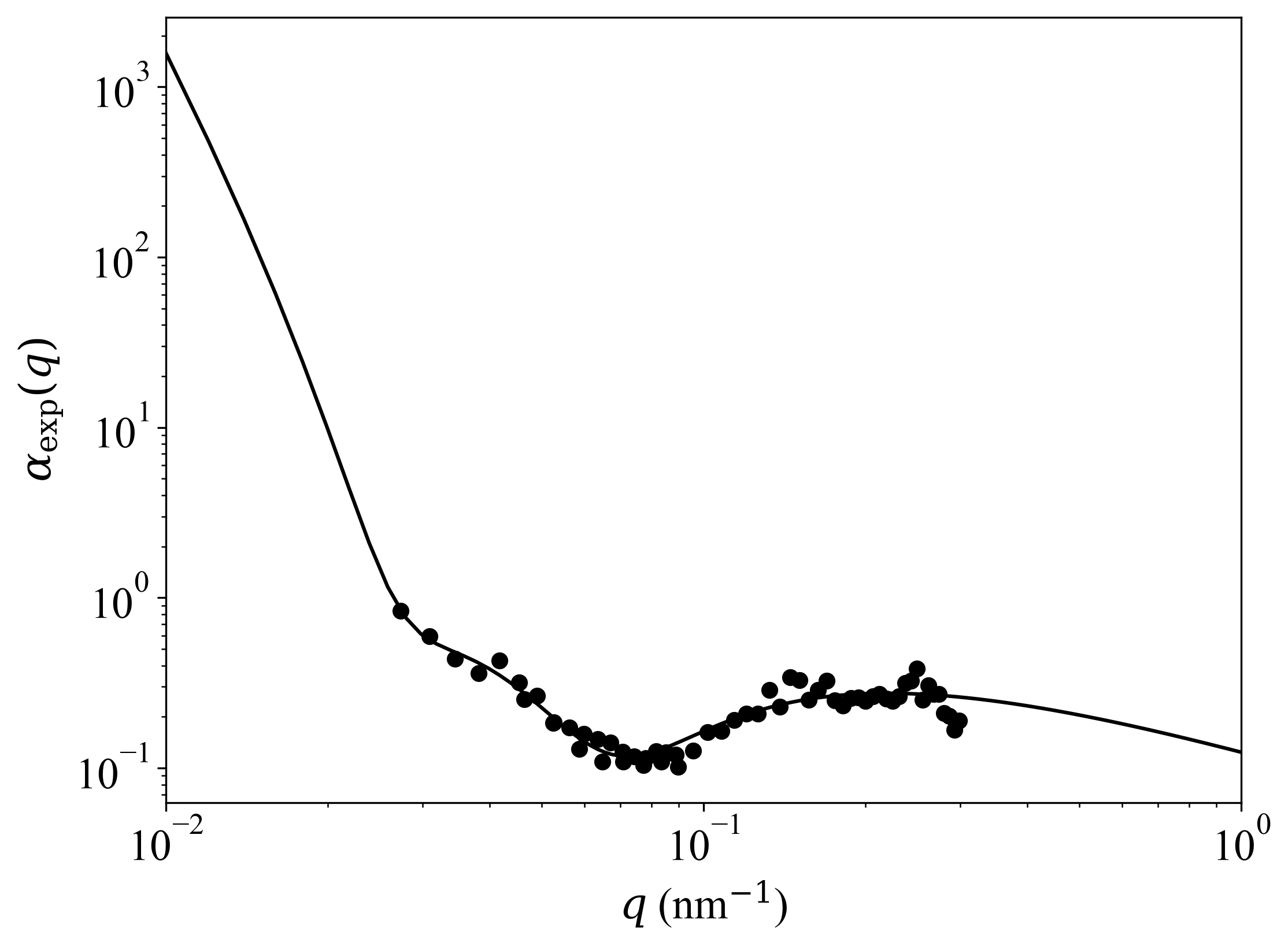}}
\caption{($\bullet$)~Experimental ratio $\alpha_{\mathrm{exp}}(q)$ of nuclear to magnetic scattering of the two-phase alloy NANOPERM~\cite{michels2012prb2} ($\mathbf{k}_0 \perp \mathbf{H}_0$; $\mu_0 H_0 = 1.27 \, \mathrm{T}$; log-log plot). Solid line:~power-law fit to parametrize the experimental data [equation~(\ref{alphafit})]. The fit has been restricted to the interval $0.03 \, \mathrm{nm}^{-1} < q < 0.3 \, \mathrm{nm}^{-1}$, but the fit function is displayed for $0.01 \, \mathrm{nm}^{-1} < q < 1.0 \, \mathrm{nm}^{-1}$.}
\label{fig4}
\end{figure}

\begin{figure}[tb!]
\centering
\resizebox{0.70\columnwidth}{!}{\includegraphics{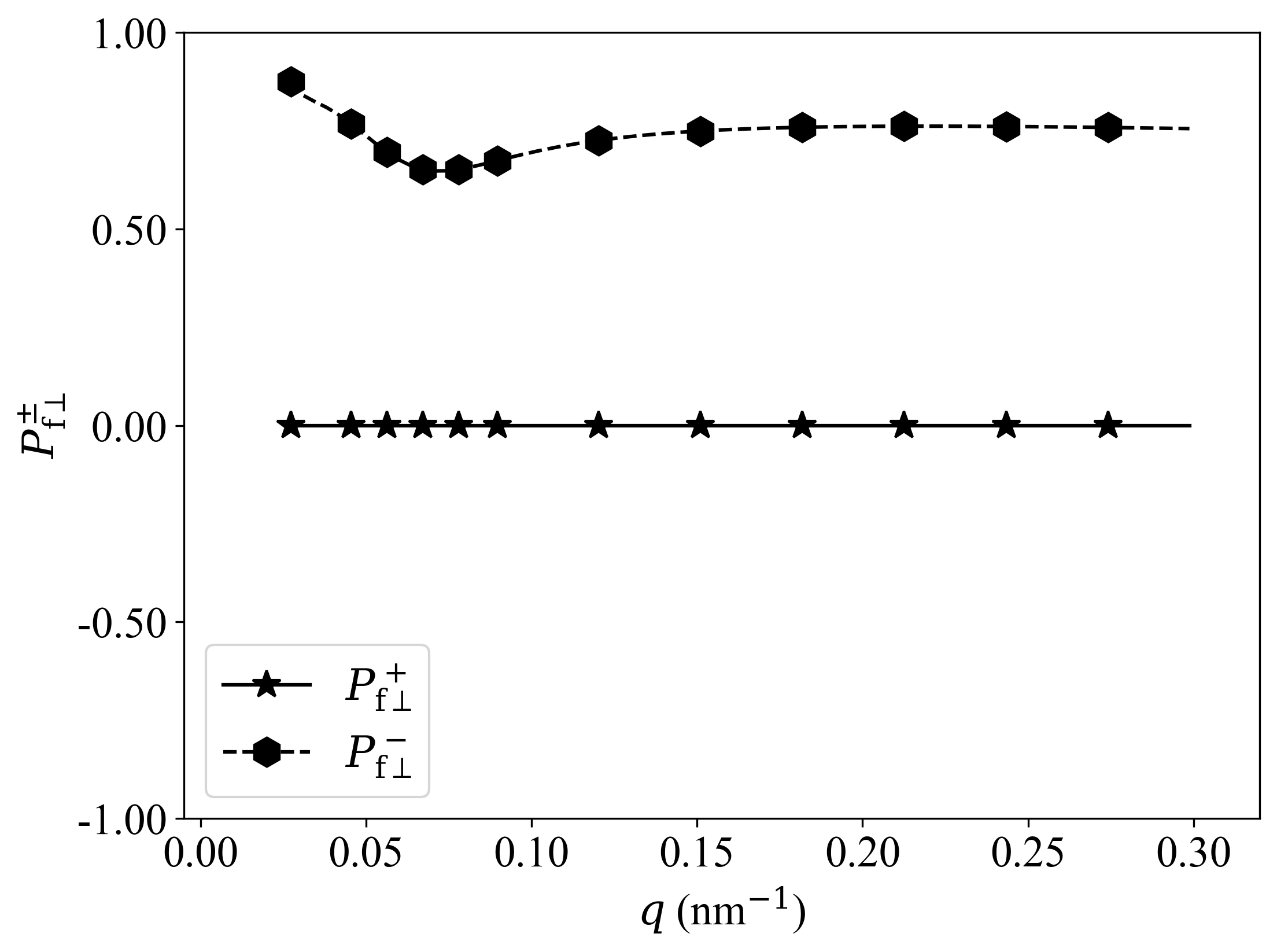}}
\caption{Plot of $P_{\mathrm{f}\perp}^{+}(q)$ (solid line) and $P_{\mathrm{f}\perp}^{-}(q)$ (dashed line) of NANOPERM using equations~(\ref{psataziplus}) and (\ref{psataziminus}) with $\alpha_{\mathrm{exp}}(q)$ given by equation~(\ref{alphafit}) and $0.03 \, \mathrm{nm}^{-1} < q < 0.3 \, \mathrm{nm}^{-1}$.}
\label{fig5}
\end{figure}

Fig.~\ref{fig4} shows the experimentally determined ratio $\alpha_{\mathrm{exp}}(q)$~\cite{michels2012prb2} of nuclear to magnetic scattering of the two-phase alloy NANOPERM. Within the experimental $q$-range of $0.03 \, \mathrm{nm}^{-1} < q < 0.3 \, \mathrm{nm}^{-1}$, these data for $\alpha_{\mathrm{exp}}(q)$ have been fitted by a power-law in $1/q$ to obtain the functions $P_{\mathrm{f}\perp}^{\pm}(q)$, which are depicted in Fig.~\ref{fig5}. The used fit function for $\alpha_{\mathrm{exp}}(q)$ is:
\begin{eqnarray}
\label{alphafit}
\alpha_{\mathrm{exp}}(q) = \frac{0.14853}{q} - \frac{0.0264491}{q^2} + \frac{0.00176887}{q^3} \nonumber \\ - \frac{4.95094 \times 10^{-5}}{q^4} + \frac{5.01767 \times 10^{-7}}{q^5} .
\end{eqnarray}
This expression will be used in the analysis of the experimental data (see section~\ref{results}).

\subsection{Nonsaturated State}

Appendix~\ref{appb} features some theoretical results for $P_{\mathrm{f}\perp}^{\pm}(\mathbf{q})$ for various combinations of the magnetic interaction parameters (applied magnetic field, ratio of $\widetilde{H}_{\mathrm{p}}^2$ to $\widetilde{M}^2_z$, $\chi(\mathbf{q}) \neq 0$). For a statistically isotropic ferromagnet, the two-dimensional distribution of the polarization of the scattered neutrons is isotropic ($\theta$-independent) for the longitudinal scattering geometry ($\mathbf{k}_0 \parallel \mathbf{H}_0$), as are the corresponding SANS cross sections. This is in contrast to the $P_{\mathrm{f}\perp}^{\pm}(\mathbf{q})$ for the transversal geometry ($\mathbf{k}_0 \perp \mathbf{H}_0$), which are highly anisotropic. In the following, we will use the theoretical expressions for $P_{\mathrm{f}\perp}^{\pm}(\mathbf{q})$ to analyze experimental data on the soft magnetic two-phase nanocrystalline alloy NANOPERM.

\section{Experimental Details}
\label{expdet}

The polarized neutron experiment was carried out at room temperature at the instrument D22 at the Institut Laue-Langevin, Grenoble, France. Incident neutrons with a mean wavelength of $\lambda = 8 \, \mathrm{{\AA}}$ and a wavelength broadening of $\Delta \lambda / \lambda = 10 \, \%$ (FWHM) were selected by means of a velocity selector. The beam was polarized using a $1.2 \, \mathrm{m}$-long remanent Fe-Si supermirror transmission polarizer ($m = 3.6$), which was installed immediately after the velocity selector. A radio-frequency (rf) spin flipper, installed close to the sample position, allowed us to reverse the initial neutron polarization. The external magnetic field (provided by an electromagnet) was applied perpendicular to the wave vector $\mathbf{k}_0$ of the incident neutrons (compare Fig.~\ref{fig1}). Measurement of the four partial POLARIS cross sections $d \Sigma^{++} / d \Omega$, $d \Sigma^{--} / d \Omega$, $d \Sigma^{+-} / d \Omega$, and $d \Sigma^{-+} / d \Omega$ was accomplished through a polarized $^3$He spin-filter cell, which was installed inside the detector housing, about $1 \, \mathrm{m}$ away from the sample position. The polarization between polarizer, rf flipper, and $^3$He filter was maintained by means of magnetic guide fields of the order of $1 \, \mathrm{mT}$. The efficiencies of the polarizer, spin flipper, and $^3$He analyzer were, respectively, $90 \, \%$, $99 \, \%$, and $87.5 \, \%$. The scattered neutrons were detected by a multitube detector which consists of $128 \times 128$ pixels with a resolution of $8 \times 8 \, \mathrm{mm}$. Neutron-data reduction, including corrections for background scattering and spin leakage~\cite{wildes06}, was performed using the GRASP~\cite{graspurl} and BerSANS~\cite{keider02,keiderling08} software packages.

The sample under study was a two-phase magnetic nanocomposite from the NANO\-PERM family of alloys~\cite{suzuki06} with a nominal composition of $(\mathrm{Fe}_{0.985}\mathrm{Co}_{0.015})_{90}\mathrm{Zr}_7\mathrm{B}_3$~\cite{suzuki1994,suzuki2007}. The alloy was prepared by melt spinning, followed by a subsequent annealing treatment for $1 \, \mathrm{h}$ at $883 \, \mathrm{K}$, which resulted in the precipitation of bcc iron nanoparticles in an amorphous magnetic matrix. The average iron particle size of $D = 15 \pm 2 \, \mathrm{nm}$ was determined by the analysis of wide-angle x-ray diffraction data. The crystalline particle volume fraction is about $65 \, \%$ and the saturation magnetization of the alloy amounts to $\mu_0 M_0 = 1.64 \, \mathrm{T}$. The exchange-stiffness constant $A = (4.7 \pm 0.9) \times 10^{-12} \, \mathrm{J/m}$ has previously been determined by the analysis of the field-dependent unpolarized SANS cross section~\cite{honecker2013}. For the SANS experiments, several circular discs with a diameter of $10 \, \mathrm{mm}$ and a thickness of about $20 \, \mu\mathrm{m}$ were stacked and mounted on a Cd aperture (for further details see Refs.~\onlinecite{michels2012prb2,honecker2013}).

\section{Experimental Results and Discussion}
\label{results}

\begin{figure}
\centering
\resizebox{1.0\columnwidth}{!}{\includegraphics{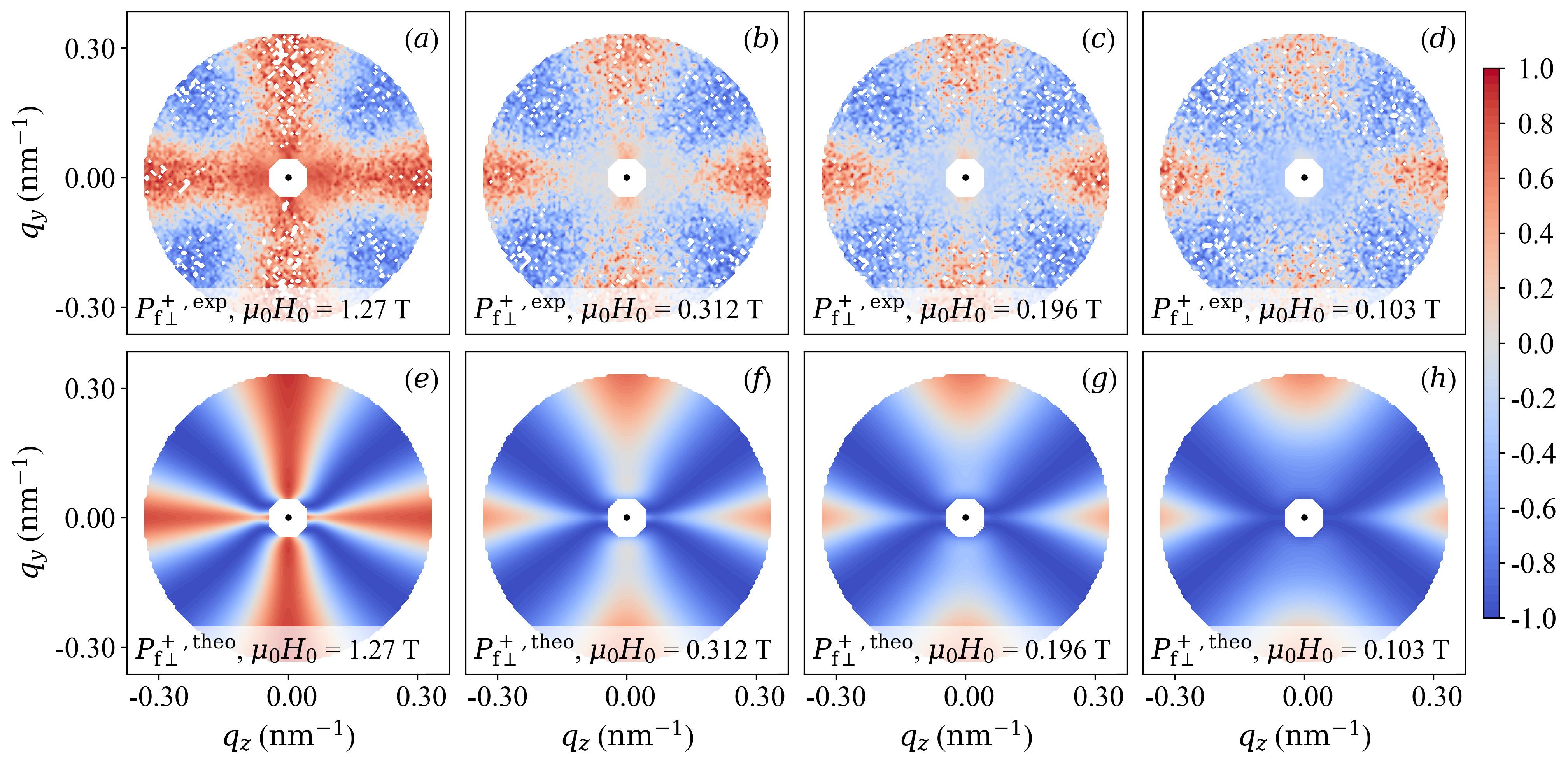}}
\caption{Qualitative comparison between experiment and theory. (\textit{a})$-$(\textit{d}) Two-dimensional experimental polarization $P_{\mathrm{f}\perp}^{+}(\mathbf{q})$ of the scattered neutrons of NANOPERM [$(\mathrm{Fe}_{0.985}\mathrm{Co}_{0.015})_{90}\mathrm{Zr}_7\mathrm{B}_3$] at a series of applied magnetic fields (see insets). $\mathbf{H}_0$ is horizontal in the plane. The range of momentum transfers is restricted to $q \lesssim 0.33 \, \mathrm{nm}^{-1}$. (\textit{e})$-$(\textit{h}) Prediction by the analytical micromagnetic theory (no free parameters) using the experimental ratio $\alpha_{\mathrm{exp}}(q)$ [equation~(\ref{alphafit})] and the structural ($\xi_{\mathrm{M}} = \xi_{\mathrm{H}} = D/2 = 7.5 \, \mathrm{nm}$) and magnetic ($A, M_0$) interaction parameters of NANOPERM (see text and Refs.~\onlinecite{michels2012prb2,honecker2013}). The central white octagons mark the position of the beam stop.}
\label{fig6}
\end{figure}

\begin{figure}
\centering
\resizebox{1.0\columnwidth}{!}{\includegraphics{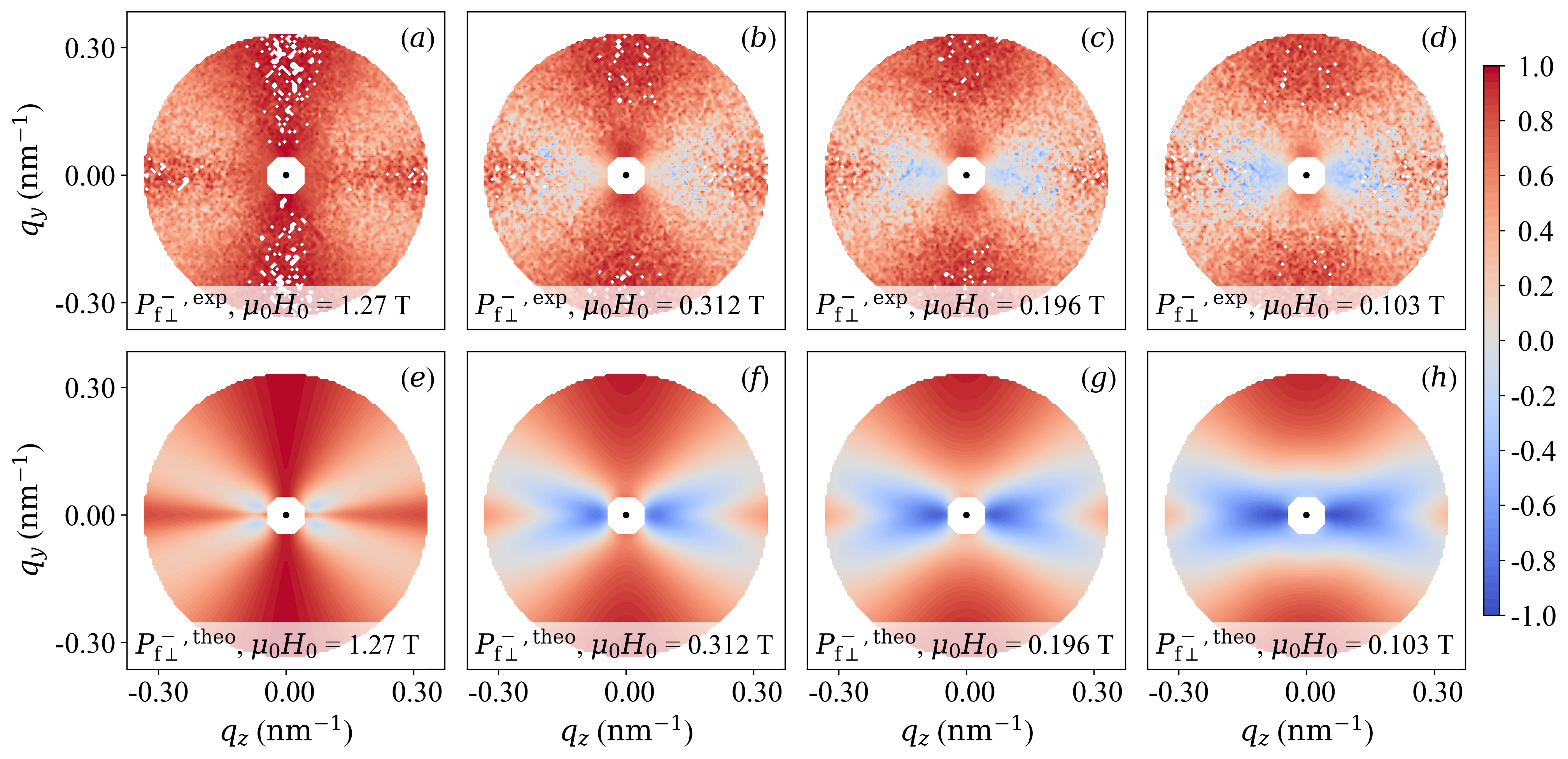}}
\caption{Similar to Fig.~\ref{fig6}, but for $P_{\mathrm{f}\perp}^{-}(\mathbf{q})$.}
\label{fig7}
\end{figure}

The two-dimensional experimental distribution of the polarization of NANOPERM is depicted in Figs.~\ref{fig6} ($P_{\mathrm{f}\perp}^{+}$) and \ref{fig7} ($P_{\mathrm{f}\perp}^{-}$) at selected field values together with a \textit{qualitative} comparison to the simulated polarization based on the micromagnetic SANS theory (see Refs.~\onlinecite{michels2012prb2} for some selected spin-resolved SANS cross sections). The theory uses as input values the experimental ratio $\alpha_{\mathrm{exp}}(q)$ [equation~(\ref{alphafit})] and the structural ($\xi_{\mathrm{M}} = \xi_{\mathrm{H}} = D/2 = 7.5 \, \mathrm{nm}$) and magnetic ($A, M_0$) interaction parameters. In agreement with the previous micromagnetic SANS data analysis of this sample~\cite{michels2012prb2,honecker2013}, we have set the ratio $A_{\mathrm{H}} / A_{\mathrm{M}} = 0.2$. We also assumed that both spin-flip channels are equal, \textit{i.e.}\ $d \Sigma^{+-} / d \Omega = d \Sigma^{-+} / d \Omega$, a constraint that was already imposed during the spin-leakage correction. The overall qualitative agreement between experiment and theory (no free parameters) is evident, although the angular anisotropy of the data does not exhibit a large variation with field. Only at the smallest momentum-transfers can one notice a change in the anisotropy with decreasing field (in particular in $P_{\mathrm{f}\perp}^{-}$), which is related to the emerging spin-misalignment scattering; compare \textit{e.g.}\ scattering terms $\propto |\widetilde{M}_y|^2 \cos^4\theta$ and $\propto CT_{yz} \sin\theta \cos\theta$ in equations~(\ref{sfperp}) and (\ref{sanspolperp}). We also note the existence of (seemingly isotropic) scattering contributions at small $q \lesssim 0.1 \, \mathrm{nm}^{-1}$ (especially at $1.27 \, \mathrm{T}$), which are likely due to large-scale structures that are not contained in the micromagnetic theory [compare Fig.~\ref{fig6}(a) and (e) and Fig.~\ref{fig7}(a) and (e)].


\begin{figure}
\centering
\resizebox{0.90\columnwidth}{!}{\includegraphics{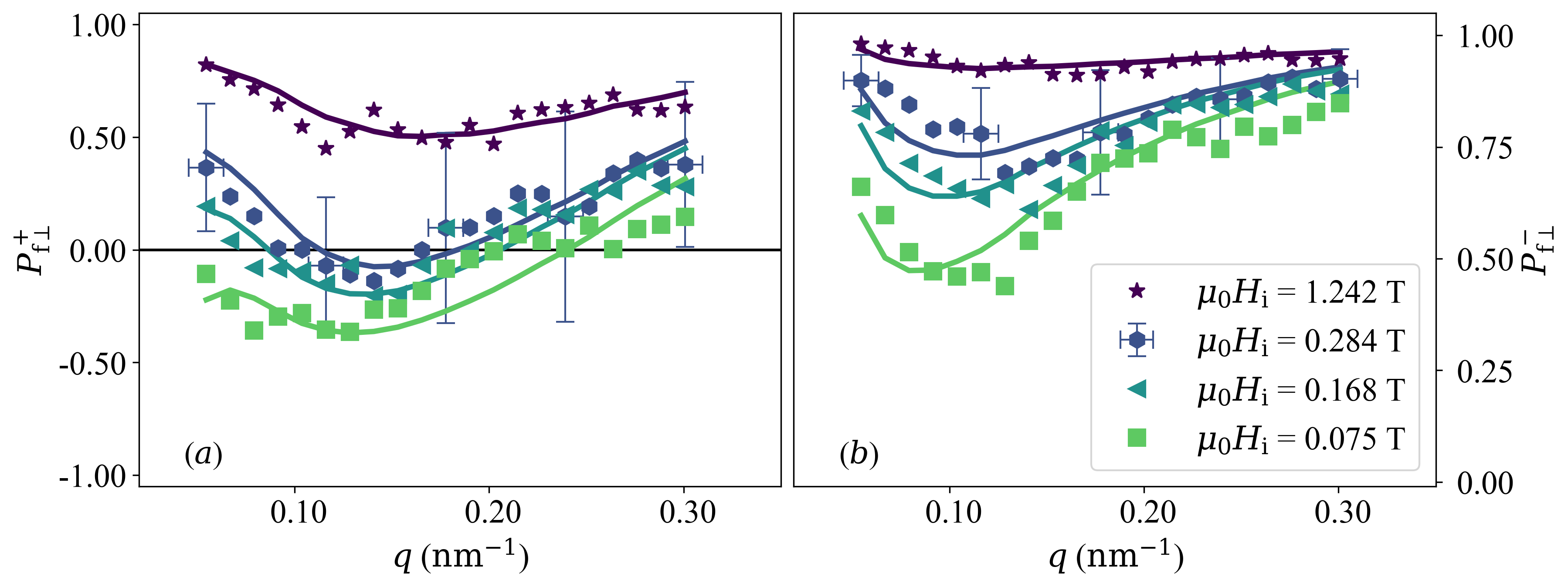}}
\caption{(Data points)~Experimental polarizations $P_{\mathrm{f}\perp}^{+}(q, \theta = 90^{\circ})$~(\textit{a}) and $P_{\mathrm{f}\perp}^{-}(q, \theta = 90^{\circ})$~(\textit{b}) of the scattered neutrons of NANOPERM [$(\mathrm{Fe}_{0.985}\mathrm{Co}_{0.015})_{90}\mathrm{Zr}_7\mathrm{B}_3$] at a series of internal magnetic fields (see inset). For the clarity of presentation, error bars are only shown for one field. (Solid lines)~Prediction by the analytical micromagnetic theory [equations~(\ref{polgen1}) and (\ref{polgen2})] using the ratio $\alpha_{\mathrm{exp}}(q)$ [equation~(\ref{alphafit})]. Note the different scales on the ordinates in (\textit{a}) and (\textit{b}).}
\label{fig8}
\end{figure}

Due to the relatively large statistical noise in the two-dimensional $P_{\mathrm{f}\perp}^{\pm}$ maps we did not fit the experimental data directly to the theoretical expressions. Therefore, in the following, we consider one-dimensional experimental polarization data, which were obtained by averaging the two-dimensional polarized SANS cross sections over $\pm 8^{\circ}$ along the vertical direction ($\theta = 90^{\circ}$). These averages were used in equations~(\ref{finalpolarizationcrosssectionplus}) and (\ref{finalpolarizationcrosssectionminus}) to obtain the $P_{\mathrm{f}\perp}^{\pm}(q)$. The resulting data in Fig.~\ref{fig8} were then fitted using the general equations~(\ref{polgen1}) and (\ref{polgen2}) (also averaged over $\pm 8^{\circ}$ along $\theta = 90^{\circ}$). Adjustable parameters are the amplitudes (scaling parameters) $A_{\mathrm{M}}$ and $A_{\mathrm{H}}$ of, respectively, $\widetilde{M}^2_z$ and $\widetilde{H}^2_{\mathrm{p}}$ as well as the corresponding correlation lengths $\xi_{\mathrm{M}}$ and $\xi_{\mathrm{H}}$ [compare equations~(\ref{mzsquaredmodel}) and (\ref{hanisquaredmodel})]. The field-dependent micromagnetic exchange length $l_{\mathrm{H}}$, which is contained in the dimensionless function $p(q, H_{\mathrm{i}})$ [compare equations~(\ref{pdef})$-$(\ref{lhdef})], is computed at each field using the materials parameters $A$ and $M_0$; $A$ is treated here as an additional adjustable parameter. For $\alpha(q)$ we used equation~(\ref{alphafit}), and the DMI has been ignored in the data analysis ($l_{\mathrm{D}} = 0$). Since the $P_{\mathrm{f}\perp}^{\pm}$ differ only by the $\widetilde{N} \widetilde{M}_y$ and $\widetilde{N} \widetilde{M}_z$~interference terms, we have fitted the $P_{\mathrm{f}\perp}^{\pm}(q)$~data corresponding to the same field simultaneously. The applied field $H_0$ has been corrected for demagnetizing effects.

\begin{figure}
\centering
\resizebox{0.65\columnwidth}{!}{\includegraphics{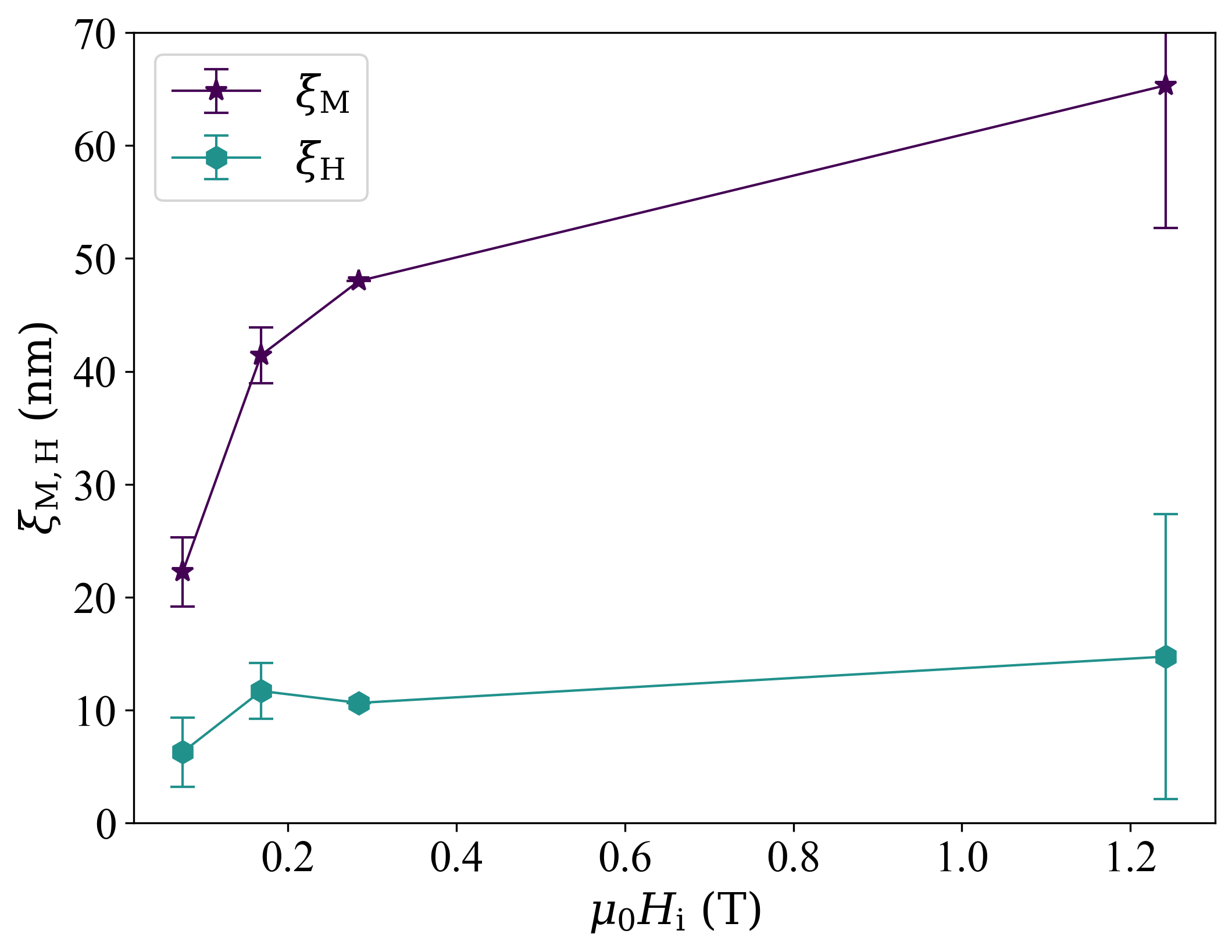}}
\caption{Resulting best-fit values for the correlation lengths $\xi_{\mathrm{M}}$ and $\xi_{\mathrm{H}}$ (see inset). Lines are a guide to the eyes.}
\label{fig9}
\end{figure}

The fits in Fig.~\ref{fig8} (solid lines) provide a reasonable description of the experimental data. The obtained values for $\xi_{\mathrm{M}}$ and $\xi_{\mathrm{H}}$ are shown in Fig.~\ref{fig9}; $\xi_{\mathrm{H}} \cong 6$$-$$15 \, \mathrm{nm}$ is at all fields consistently of the order of the particle size, while $\xi_{\mathrm{M}}$ takes on larger values between about $22$$-$$65 \, \mathrm{nm}$. For the exchange-stiffness constant, we obtain (from the four local fits) best-fit values that range between $A = (4.8$$-$$9.7) \times 10^{-12} \, \mathrm{J/m}$. These values agree very well with data in the literature~\cite{honecker2013,mathiasiucrj2021}.

Clearly, more experiments are needed in order to further scrutinize the predictions of the present micromagnetic theory for the uniaxial polarization analysis of bulk ferromagnets. In this respect, the development of computational tools to directly analyze the two-dimensional polarization maps using different form-factor and structure-factor expressions for $\widetilde{M}^2_z$ and $\widetilde{H}^2_{\mathrm{p}}$, and possibly the inclusion of a particle-size distribution function, would be desirable. Likewise, SANS measurements at a preferably saturating magnetic field are necessary to determine the nuclear SANS cross section, \textit{e.g.}\ via a horizontal average of the non-spin-flip SANS cross section.

\section{Summary and Conclusions}
\label{summary}

We have provided a micromagnetic theory for the uniaxial polarization of the scattered neutrons of bulk ferromagnets, as it can be measured by means of the small-angle neutron scattering (SANS) method. The theoretical expressions contain the effects of an isotropic exchange interaction, the Dzyaloshinskii--Moriya interaction, magnetic anisotropy, magnetodipolar interaction, and an external magnetic field. The theory has been employed to analyze experimental data on a soft magnetic nanocrystalline alloy; it may provide information on the magnetic interactions (exchange and DMI constants) and on the spatial structures of the magnetic anisotropy and magnetostatic fields. Given that uniaxial polarization analysis is becoming more and more available on SANS instruments worldwide and in view of the recent seminal progress made regarding several techniques which exploit the neutron polarization degree of freedom to characterize large-scale magnetic structures (SESANS, DFI, SEMSANS), we believe that the results of this paper open up a new avenue for magnetic neutron data analysis on mesoscopic magnetic systems. The presented micromagnetic SANS framework forms the basis for all of these new and promising neutron techniques.

\acknowledgements

Andreas Michels and Artem Malyeyev thank the National Research Fund of Luxembourg for financial support (AFR Grant No.~12417141). This work is partially based on experiments performed at the Institut Laue-Langevin, Grenoble, France.

\bibliography{BIB}
\bibliographystyle{apsrev4-2}

\appendix
\section{Non-spin-flip, spin-flip, and SANSPOL cross sections}
\label{appa}

In this Appendix, we display the expressions for the polarized SANS cross sections in terms of the Fourier components $\widetilde{M}_{x,y,z}(\mathbf{q})$ of the magnetization. The two non-spin-flip and the two spin-flip POLARIS cross sections carry, respectively, the superscripts $+ +$ and $- -$ and $+ -$ and $- +$, and the subscripts $\perp$ and $\parallel$ refer to the respective scattering geometry (compare Fig.~\ref{fig1})~\cite{michels2014review}:
\begin{eqnarray}
\label{nsfperp}
\frac{d \Sigma^{\pm \pm}_{\perp}}{d \Omega} = K \left( b_{\mathrm{H}}^{-2} |\widetilde{N}|^2 + |\widetilde{M}_y|^2 \sin^2\theta \cos^2\theta + |\widetilde{M}_z|^2 \sin^4\theta \right. \nonumber \\ \left. - CT_{yz} \sin^3\theta \cos\theta \mp b_{\mathrm{H}}^{-1} CT_{\widetilde{N} \widetilde{M}_z} \sin^2\theta \right. \nonumber \\ \left. \pm b_{\mathrm{H}}^{-1} CT_{\widetilde{N} \widetilde{M}_y} \sin\theta \cos\theta \right) , \nonumber \\
\end{eqnarray}
\begin{eqnarray}
\label{nsfpara}
\frac{d \Sigma^{\pm \pm}_{\parallel}}{d \Omega} = K \left( b_{\mathrm{H}}^{-2} |\widetilde{N}|^2 + |\widetilde{M}_z|^2 \mp b_{\mathrm{H}}^{-1} CT_{\widetilde{N} \widetilde{M}_z} \right) , \nonumber \\
\end{eqnarray}
\begin{eqnarray}
\label{sfperp}
\frac{d \Sigma^{\pm \mp}_{\perp}}{d \Omega} = K \left( |\widetilde{M}_x|^2 + |\widetilde{M}_y|^2 \cos^4\theta + |\widetilde{M}_z|^2 \sin^2\theta \cos^2\theta \right. \nonumber \\ \left. - CT_{yz} \sin\theta \cos^3\theta \mp i \chi \right) \nonumber , \\
\end{eqnarray}
\begin{eqnarray}
\label{sfpara}
\frac{d \Sigma^{\pm \mp}_{\parallel}}{d \Omega} = K \left( |\widetilde{M}_x|^2 \sin^2\theta + |\widetilde{M}_y|^2 \cos^2\theta \right. \nonumber \\ \left. - CT_{xy} \sin\theta \cos\theta \right) ,
\end{eqnarray}
where $K = 8 \pi^3 V^{-1} b_{\mathrm{H}}^2$, and the chiral function $\chi(\mathbf{q})$ is given by:
\begin{eqnarray}
\label{chiral}
\chi = \left( \widetilde{M}_x \widetilde{M}_y^{\ast} - \widetilde{M}_x^{\ast} \widetilde{M}_y \right) \cos^2\theta \nonumber \\ - \left( \widetilde{M}_x \widetilde{M}_z^{\ast} - \widetilde{M}_x^{\ast} \widetilde{M}_z \right) \sin\theta \cos\theta .
\end{eqnarray}
Note that $\chi(\mathbf{q}) = 0$ for $\mathbf{k}_0 \parallel \mathbf{H}_0$. The two SANSPOL cross sections $\frac{d \Sigma^{+}}{d \Omega} = \frac{d \Sigma^{++}}{d \Omega} + \frac{d \Sigma^{+-}}{d \Omega}$ and $\frac{d \Sigma^{-}}{d \Omega} = \frac{d \Sigma^{--}}{d \Omega} + \frac{d \Sigma^{-+}}{d \Omega}$ read:
\begin{eqnarray}
\label{sanspolperp}
\frac{d \Sigma^{\pm}_{\perp}}{d \Omega} = K \left( b_{\mathrm{H}}^{-2} |\widetilde{N}|^2 + |\widetilde{M}_x|^2 + |\widetilde{M}_y|^2 \cos^2\theta \right. \nonumber \\ \left. + |\widetilde{M}_z|^2 \sin^2\theta - CT_{yz} \sin\theta \cos\theta \right. \nonumber \\ \left. \mp b_{\mathrm{H}}^{-1} CT_{\widetilde{N} \widetilde{M}_z} \sin^2\theta \pm b_{\mathrm{H}}^{-1} CT_{\widetilde{N} \widetilde{M}_y} \sin\theta \cos\theta \right. \nonumber \\ \left. \mp i \chi \right) ,
\end{eqnarray}
\begin{eqnarray}
\label{sanspolpara}
\frac{d \Sigma^{\pm}_{\parallel}}{d \Omega} = K \left( b_{\mathrm{H}}^{-2} |\widetilde{N}|^2 + |\widetilde{M}_x|^2 \sin^2\theta + |\widetilde{M}_y|^2 \cos^2\theta \right. \nonumber \\ \left. + |\widetilde{M}_z|^2 - CT_{xy} \sin\theta \cos\theta \mp b_{\mathrm{H}}^{-1} CT_{\widetilde{N} \widetilde{M}_z} \right) . \nonumber \\
\end{eqnarray}
The magnetic-magnetic and nuclear-magnetic cross terms have been abbreviated as follows:
\begin{subequations}
\begin{eqnarray}
\label{eq:ctyz}
CT_{yz} &=& \widetilde{M}_y \widetilde{M}_z^{\ast} + \widetilde{M}_y^{\ast} \widetilde{M}_z , \\ 
CT_{xy} &=& \widetilde{M}_x \widetilde{M}_y^{\ast} + \widetilde{M}_x^{\ast} \widetilde{M}_y , \\
CT_{\widetilde{N} \widetilde{M}_z} &=& \widetilde{N} \widetilde{M}_z^{\ast} + \widetilde{N}^{\ast} \widetilde{M}_z , \\
CT_{\widetilde{N} \widetilde{M}_y} &=& \widetilde{N} \widetilde{M}_y^{\ast} + \widetilde{N}^{\ast} \widetilde{M}_y .
\end{eqnarray}
\end{subequations}
In actual SANSPOL and POLARIS experiments the neutron optics (polarizer, spin flipper, analyzer) do not work perfectly and polarization corrections become necessary. The incident beam polarization efficiency may be denoted by $p^+ = I^+/(I^+ + I^-)$, where $I^{\pm}$ are, respectively, the number of neutrons with spins aligned antiparallel and parallel with respect to $\mathbf{H}_0$; note that $p^+ = 1/2$ for an unpolarized beam. The efficiency of the spin flipper is $\epsilon^{\pm}$ with $\epsilon^{+} = 0$ for flipper off and $\epsilon^{-} = \epsilon \cong 1$ for flipper on. It is important to emphasize that the half-polarized SANSPOL cross sections $d \Sigma^{+} / d \Omega$ and $d \Sigma^{-} / d \Omega$ can be obtained directly and corrected for nonideal neutron polarization provided that the parameters $p^+$ and $\epsilon$ are known from reference measurements. For POLARIS, it is necessary to measure all four partial cross sections $d \Sigma^{++} / d \Omega$, $d \Sigma^{--} / d \Omega$, $d \Sigma^{+-} / d \Omega$, and $d \Sigma^{-+} / d \Omega$ in order to correct for spin leakage between the different channels~\cite{wildes06}. Such corrections can \textit{e.g.}\ be accomplished by means of the \textit{BerSANS}~\cite{keider02,keiderling08}, \textit{Pol-Corr}~\cite{krycka2012a}, and \textit{GRASP}~\cite{graspurl} software tools.

Moreover, we note that $d \Sigma^{+-} / d \Omega = d \Sigma^{-+} / d \Omega$ for many polycrystalline bulk ferromagnets~\cite{michels2010epjb}. However, in our theoretical treatment we explicitly take into account the polarization dependence of the SANSPOL and spin-flip cross sections via the chiral function $\chi(\mathbf{q})$. This is relevant \textit{e.g.}\ for systems where inversion symmetry is broken and the DMI is operative~\cite{michelsdmi2019,quan2020}.

\section{Selected results for the polarization of the scattered neutrons of bulk ferromagnets}
\label{appb}

In this Appendix we provide some selected graphical representations for the dependency of the polarization of the scattered neutrons on the magnitude and orientation of the scattering vector, the applied magnetic field, the ratio of $A_{\mathrm{M}}$ to $A_{\mathrm{H}}$, the ratio of nuclear to longitudinal magnetic scattering, and the DMI (via the exchange length $l_{\mathrm{D}}$). Only results for $\mathbf{k}_0 \perp \mathbf{H}_0$ are shown. The following materials parameters are used:~$A = 4.7 \, \mathrm{pJ/m}$; $\mu_0 M_0 = 1.64 \, \mathrm{T}$ ($l_{\mathrm{M}} \cong 2.1 \, \mathrm{nm}$); $D = 2 \, \mathrm{mJ/m^2}$ ($l_{\mathrm{D}} \cong 1.9 \, \mathrm{nm}$)~\cite{honecker2013}.

\begin{figure}
\centering
\resizebox{0.90\columnwidth}{!}{\includegraphics{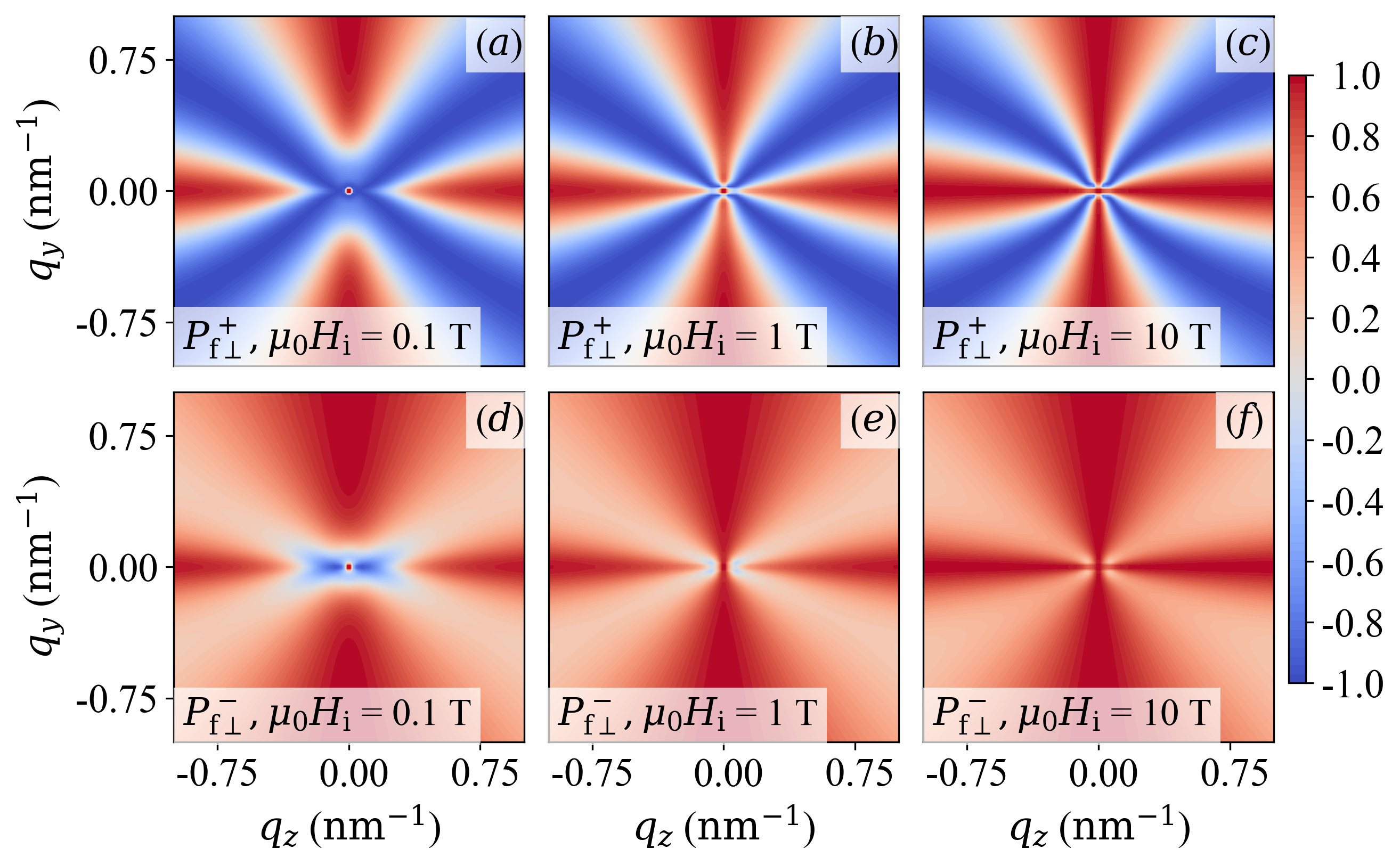}}
\caption{Plot of $P_{\mathrm{f}\perp}^{+}(q_y, q_z)$ (upper row) and $P_{\mathrm{f}\perp}^{-}(q_y, q_z)$ (lower row) for different applied magnetic fields $H_{\mathrm{i}}$ (see insets). $\alpha = \alpha(q)$ [equation~(\ref{alphafit})], $A_{\mathrm{H}} / A_{\mathrm{M}} = 0.5$, and $l_{\mathrm{D}} = 0$.}
\label{figa1}
\end{figure}

\begin{figure}
\centering
\resizebox{0.95\columnwidth}{!}{\includegraphics{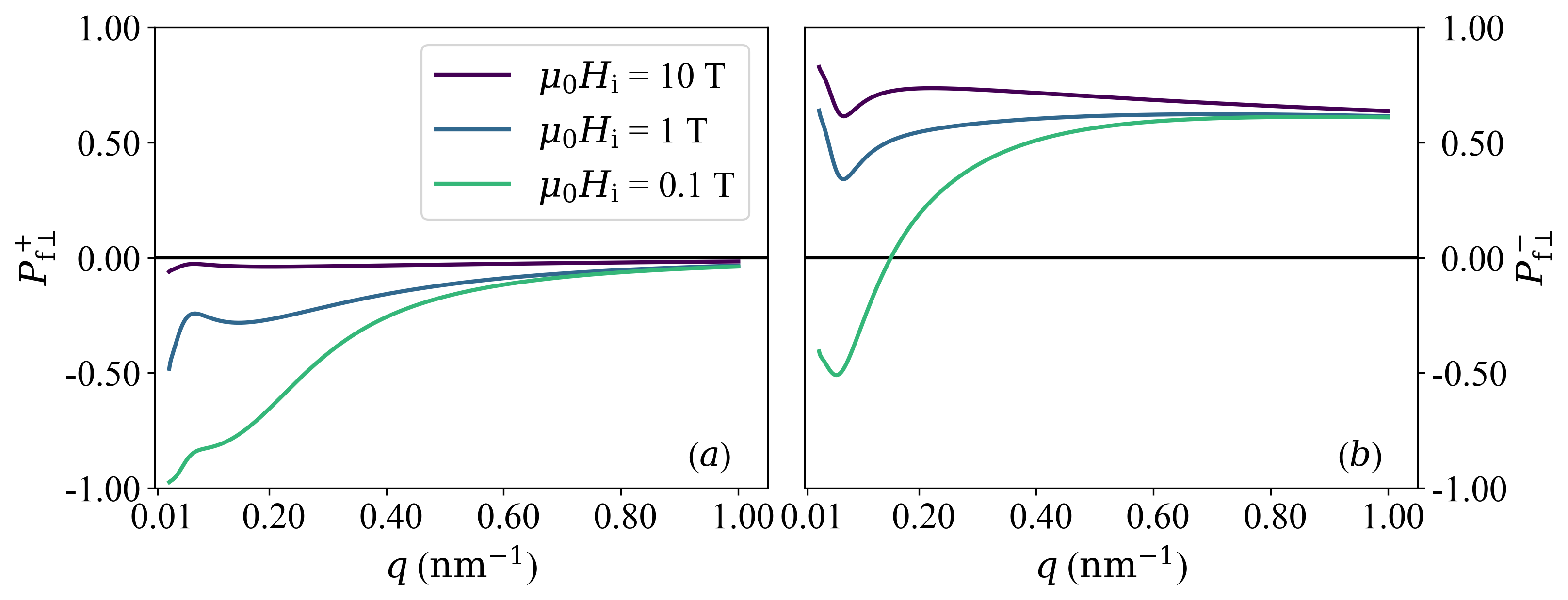}}
\caption{$2\pi$-azimuthally-averaged $P_{\mathrm{f}\perp}^{+}(q)$~(\textit{a}) and $P_{\mathrm{f}\perp}^{-}(q)$~(\textit{b}) of the data shown in Fig.~\ref{figa1}.}
\label{figa2}
\end{figure}

\begin{figure}
\centering
\resizebox{0.90\columnwidth}{!}{\includegraphics{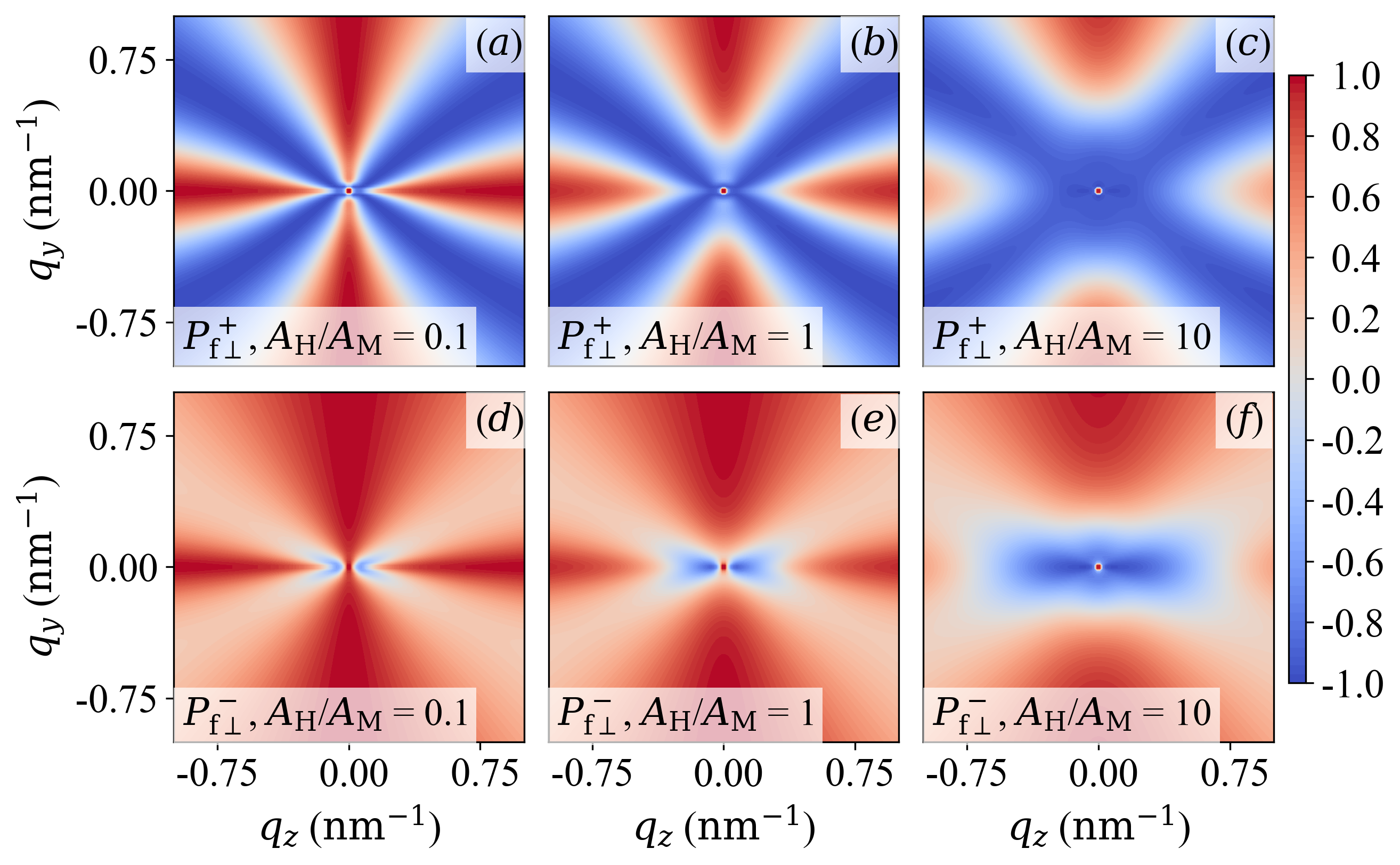}}
\caption{Plot of $P_{\mathrm{f}\perp}^{+}(q_y, q_z)$ (upper row) and $P_{\mathrm{f}\perp}^{-}(q_y, q_z)$ (lower row) for different ratios of $A_{\mathrm{H}} / A_{\mathrm{M}}$ (see insets). $\alpha = \alpha(q)$ [equation~(\ref{alphafit})], $\mu_0 H_{\mathrm{i}} = 0.3 \, \mathrm{T}$, and $l_{\mathrm{D}} = 0$.}
\label{figa3}
\end{figure}

\begin{figure}
\centering
\resizebox{0.90\columnwidth}{!}{\includegraphics{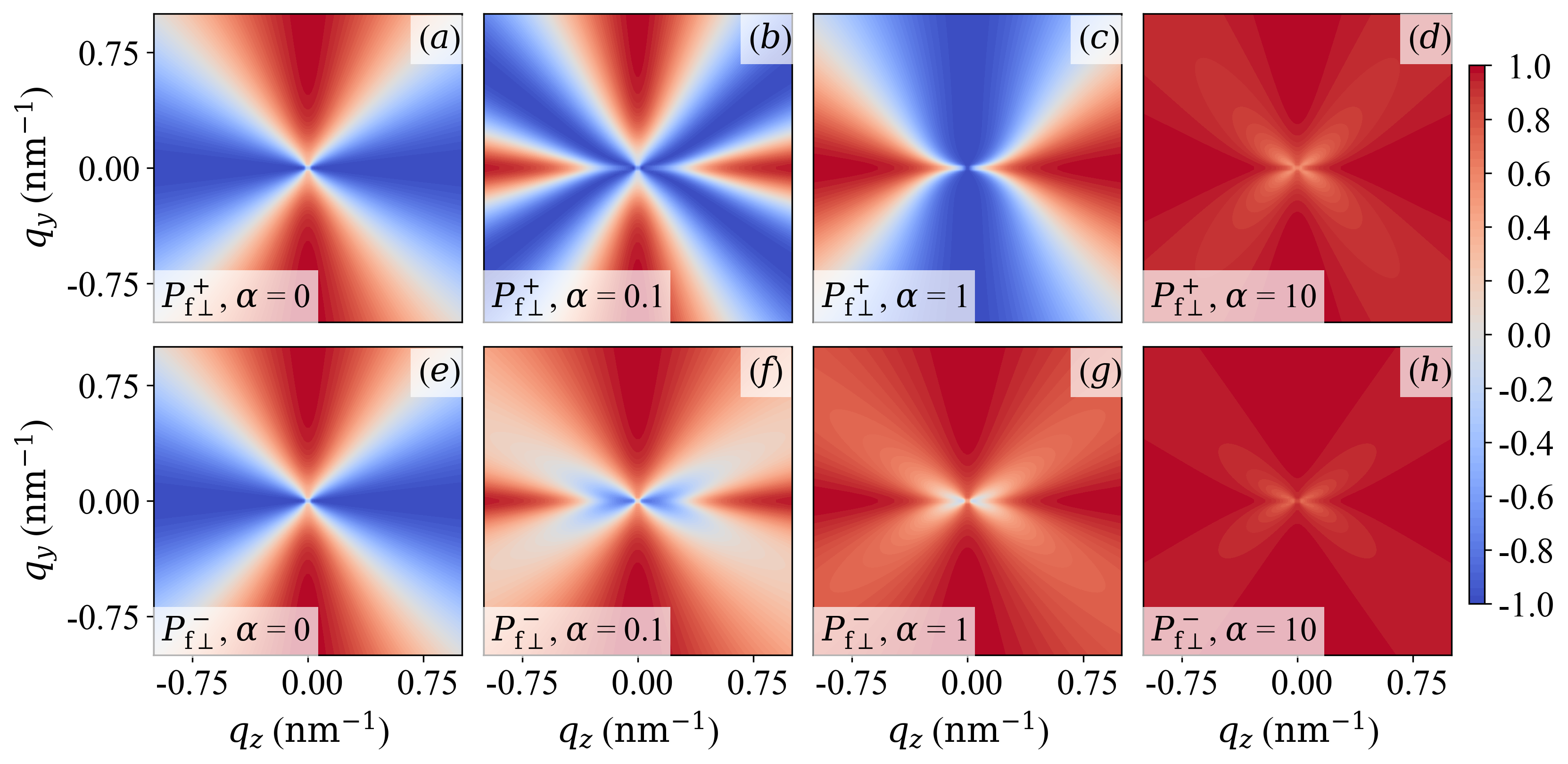}}
\caption{Plot of $P_{\mathrm{f}\perp}^{+}(q_y, q_z)$ (upper row) and $P_{\mathrm{f}\perp}^{-}(q_y, q_z)$ (lower row) for different values of $\alpha = \mathrm{constant}$ (see insets). $A_{\mathrm{H}} / A_{\mathrm{M}} = 0.2$, $\mu_0 H_{\mathrm{i}} = 0.3 \, \mathrm{T}$, and $l_{\mathrm{D}} = 0$.}
\label{figa4}
\end{figure}

\begin{figure}
\centering
\resizebox{0.90\columnwidth}{!}{\includegraphics{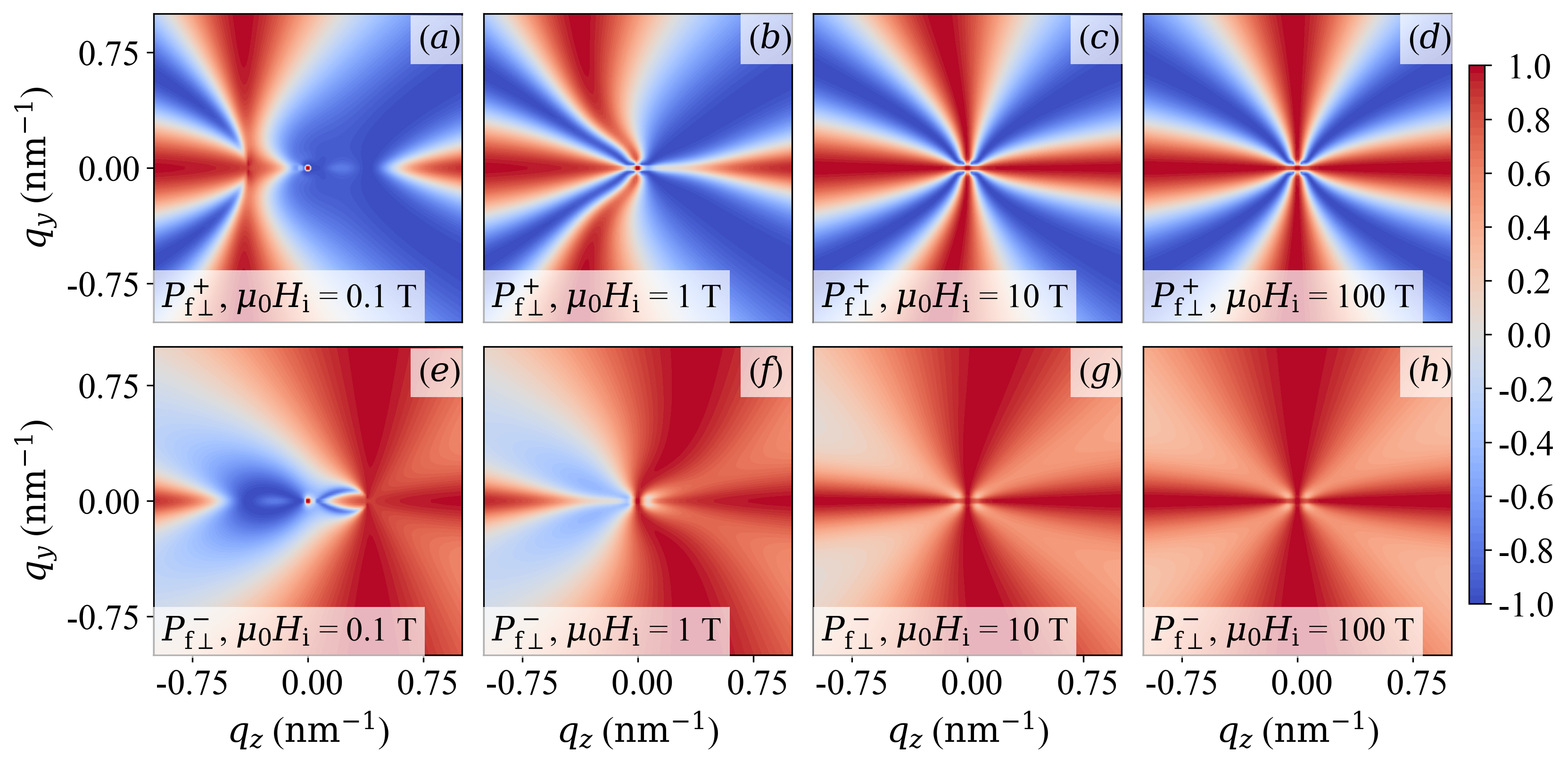}}
\caption{Effect of the DMI. Plot of $P_{\mathrm{f}\perp}^{+}(q_y, q_z)$ (upper row) and $P_{\mathrm{f}\perp}^{-}(q_y, q_z)$ (lower row) as a function of $H_{\mathrm{i}}$ (see insets). $\alpha = \alpha(q)$ [equation~(\ref{alphafit})], $A_{\mathrm{H}} / A_{\mathrm{M}} = 1$, and $l_{\mathrm{D}} = 1.9 \, \mathrm{nm}$.}
\label{figa5}
\end{figure}

\begin{figure}
\centering
\resizebox{0.60\columnwidth}{!}{\includegraphics{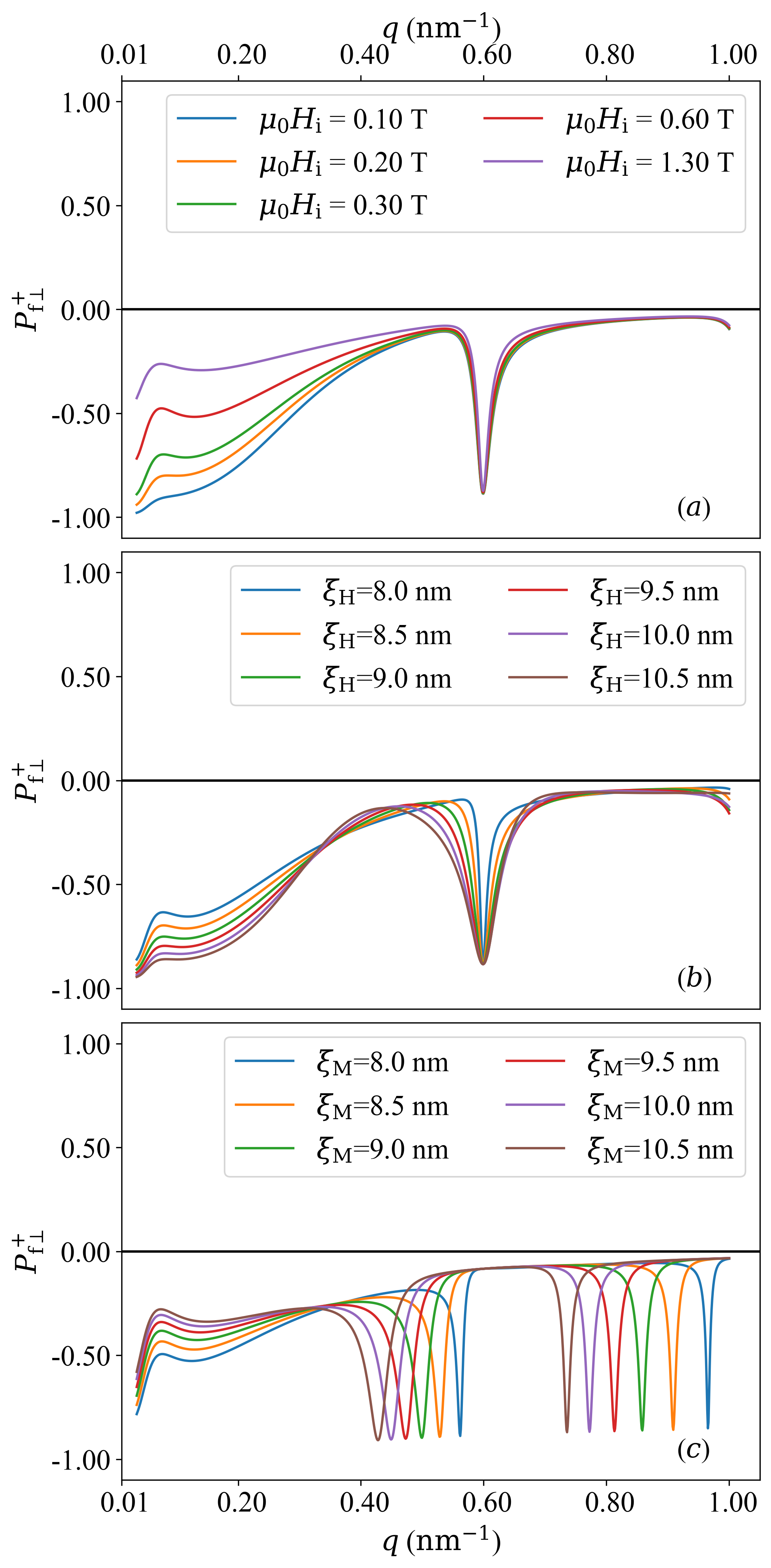}}
\caption{Results for the azimuthally-averaged $P_{\mathrm{f}\perp}^{+}(q)$ using the sphere form factor (instead of Lorentzian squared functions) for both $\widetilde{M}_z^2(q \xi_{\mathrm{M}})$ and $\widetilde{H}_{\mathrm{p}}^2(q \xi_{\mathrm{H}})$. (\textit{a})~Field dependence (see inset) of $P_{\mathrm{f}\perp}^{+}(q)$ for $\xi_{\mathrm{M}} = 7.5 \, \mathrm{nm}$ and $\xi_{\mathrm{H}} = 8.5 \, \mathrm{nm}$. (\textit{b})~$P_{\mathrm{f}\perp}^{+}(q)$ at $\mu_0 H_{\mathrm{i}} = 0.3 \, \mathrm{T}$, $\xi_{\mathrm{M}} = 7.5 \, \mathrm{nm}$, but for increasing $\xi_{\mathrm{H}}$ (see inset). (\textit{c})~$P_{\mathrm{f}\perp}^{+}(q)$ at $\mu_0 H_{\mathrm{i}} = 0.3 \, \mathrm{T}$, $\xi_{\mathrm{H}} = 7.5 \, \mathrm{nm}$, but for increasing $\xi_{\mathrm{M}}$ (see inset). $\alpha = \alpha(q)$ [equation~(\ref{alphafit})], $A_{\mathrm{H}} / A_{\mathrm{M}} = 0.2$, and $l_{\mathrm{D}} = 0$.}
\label{figa6}
\end{figure}

Fig.~\ref{figa1} shows the two-dimensional final polarization $P_{\mathrm{f}\perp}^{\pm}(q_y, q_z)$ and Fig.~\ref{figa2} depicts the corresponding $2\pi$-azimuthally-averaged data $P_{\mathrm{f}\perp}^{\pm}(q)$ for different values of the applied magnetic field $H_{\mathrm{i}}$, $\alpha = \alpha(q)$ [equation~(\ref{alphafit})], $A_{\mathrm{H}} / A_{\mathrm{M}} = 0.5$, and $l_{\mathrm{D}} = 0$. The local extrema in $P_{\mathrm{f}\perp}^{\pm}(q)$ at small $q$ are due to $\alpha_{\mathrm{exp}}(q)$; setting $\alpha=$~constant results in smooth and continuously decaying functions. Figs.~\ref{figa3} and \ref{figa4} display the polarization $P^{\pm}(\mathbf{q})$ for different ratios of $A_{\mathrm{H}} / A_{\mathrm{M}}$ (Fig.~\ref{figa3}) and for different (constant) $\alpha$-values (Fig.~\ref{figa4}) at a constant field of $\mu_0 H_{\mathrm{i}} = 0.3 \, \mathrm{T}$ and for $l_{\mathrm{D}} = 0$. Including the DMI results in asymmetric $P^{\pm}$~patterns at nonsaturating fields (see Fig.~\ref{figa5}). For the calculation of the $P^{\pm}$ according to equation~(\ref{polgen1}), we have up to now assumed Lorentzian-squared functions for $\widetilde{M}_z^2(q \xi_{\mathrm{M}})$ and $\widetilde{H}_{\mathrm{p}}^2(q \xi_{\mathrm{H}})$ [compare equations~(\ref{mzsquaredmodel})$-$(\ref{hanisquaredmodel})] with $\xi_{\mathrm{M}} = \xi_{\mathrm{H}} = D/2 = 7.5 \, \mathrm{nm}$. The effect of a hard-sphere form factor for $\widetilde{M}_z^2$ and $\widetilde{H}_{\mathrm{p}}^2$ with \textit{different} values for $\xi_{\mathrm{M}}$ and $\xi_{\mathrm{H}}$ is depicted in Fig.~\ref{figa6}. Here, peak-type features may appear in $P^{\pm}$, which might be detected in highly monodisperse particulate systems.

\end{document}